\documentclass[aps,pra,reprint,twocolumn,superscriptaddress]{revtex4-2}
\usepackage{bm}
\usepackage{amssymb} 
\usepackage{amsmath}
\usepackage[colorlinks,citecolor=blue,pdftex]{hyperref}
\usepackage{graphicx}
\usepackage{color}
\usepackage{textgreek}
\usepackage{hyperref}
\usepackage{xcolor}
\usepackage{ulem}

\begin{document}


\title{Nonclassicality in Two-Mode Stabilized Squeezed Coherent State: Quantum-to-Classical transition}

\author{Chanseul Lee}%
\affiliation{%
	Department of Physics, Korea University, 145 Anam-ro, Seongbuk-gu, Seoul 02841, Republic of Korea}
\author{Tai Hyun Yoon}%
\email{thyoon@korea.ac.kr (THY)}
\affiliation{%
	Department of Physics, Korea University, 145 Anam-ro, Seongbuk-gu, Seoul 02841, Republic of Korea}
\affiliation{%
	Center for Molecular Spectroscopy and Dynamics, Institute for Basic Science (IBS), Korea University, 145 Anam-ro, Seongbuk-gu, Seoul 02841, Republic of Korea}


\date{\today}

\begin{abstract}
We consider a two-mode stabilized squeezed coherent state (SSCS) of light and introduce the $\Pi_{\rm N}$ indicator, a novel measure for characterizing nonclassicality in the resulting EPR-entangled state. Unlike existing methods based on Cauchy-Schwarz or Murihead inequalities, $\Pi_{\rm N}$ leverages analytical solutions to the quantum Langevin equations to directly analyze nonclassicality arising from key processes like bichromatic injection, frequency conversion, and parametric down-conversion (both spontaneous and stimulated). This approach not only identifies the optimal phase for maximum nonclassicality but also reveals two new phenomena: first, both intra-cavity and extra-cavity fields exhibit the same degree of nonclassicality, and second, balanced seeding in phase-mismatched configurations induces nonclassicality across a broad range of squeezing and seeding parameters. Our work deepens the understanding of the intricate dependence of nonclassicality on system parameters in the context of SSCS, paving the way for investigations into the quantum-to-classical transition in entangled systems. The potential of $\Pi_{\rm N}$ holds significant promise for advancements in quantum optics and information science.
\end{abstract}

\maketitle

\section{Introduction \label{sec:Int}}
Quantitative nonclassicality measures, crucial to quantum information science, enable the implementation of quantum information protocols \cite{Furusawa98,Braunstein05,Menicucci06,Yadin18,Kwon19} using controllable system parameters of quantum states. These protocols facilitate high-precision measurements \cite{Oelker16,Ge20,Korobko23,Otabe24}, surpassing the classical limitations imposed by quantum projection noise. Ultimately, these measures can help reveal the mechanisms behind the quantum-to-classical transition \cite{Zurek03,Zavatta04,Schlosshauer07}.

In quantum optics, for single-mode light, nonclassicality measures have been investigated based on various concepts, including quantum Fisher information \cite{Rivas10,Tan17,Kwon19}, entanglement potential \cite{Asboth05}, nonclassical distance \cite{Hillery87}, Bures distance \cite{Marian02}, nonclassical depth \cite{Lee91}, and nonclassicality with positive semidefinite observables \cite{Gehrke12}. In the case of multi-mode light, different approaches are used. For example, nonclassicality measures are often parameterized by quantum contextuality \cite{Cabello13,Shafiee24}, logarithmic negativity of the covariance matrix \cite{Vogel14}, negative quasiprobability \cite{Sperling09,Kiesel11,Rahimi13,Tan20}, or violations of inequalities like the Cauchy-Schwarz and Murihead inequalities \cite{Reid86,Agarwal88,Lee90a,Li00,Marino08,Wasak14,Volovich15}.

In particular, for two-mode entangled light, violation of classical inequalities, like the Cauchy-Schwarz inequality (CSI) \cite{Reid86}, indicates the presence of strong quantum correlations between the modes (EPR-entanglement). For example, a state with a negative Glauber $P$ function \cite{Glauber63a,Titulaer65,Sudarshan63} signifies a violation of the CSI \cite{Reid86}. Introduced by Agarwal \cite{Agarwal88}, this criterion ($\Pi_{\rm A}$ in Eq.~(\ref{E23})) characterizes the quantum properties of general two-mode states and remains applicable even in the single-photon regime. It incorporates both the mean values of self and cross-correlation functions for the two-mode number operators. Similarly, Lee proposed a slightly different definition for the second-order moment (SOM) criterion ($\Pi_{\rm L}$ in Eq.~(\ref{EM})) based on the Murihead inequality, which is equivalent to the elementary arithmetic-geometric mean inequality and relies on the majorization of higher-order moments of number operators \cite{Lee90a}.

However, all-known nonclassicality measures fail to analytically quantify the contributions from individual quantum mechanisms within the quantum light source. This significantly hinders our ability to quantitatively reveal the quantum-to-classical transition as this system interacts with the environment.  

This paper investigates the applicability of existing nonclassicality measures ($\Pi_{\rm A}$ and $\Pi_{\rm L}$) to a two-mode stabilized squeezed coherent state (SSCS) generated by a seeded optical parametric oscillator (OPO) in the single-photon regime (Fig.~\ref{Fig0}). While $\Pi_{\rm A}$ and $\Pi_{\rm L}$ show the same expressions both inside and outside the OPO cavity, they fail to distinguish the separate contributions of individual system parameters. Our goal is to quantitatively understand how the quantum-to-classical transition of the SSCS depends on the strength of controllable system parameters.

To overcome limitations in existing nonclassicality measures' ability to quantify individual quantum processes, we introduce a novel quantitative nonclassicality indicator, $\Pi_{\rm N}$ 
in Eq.~(\ref{NomPi}), based on the Murihead inequality and normalized by the contribution from spontaneous parametric down-conversion (SPDC). Notably, $\Pi_{\rm N}$ separates and quantifies the contributions of four distinct SOM components.

Among these mechanisms, the frequency conversion (FC) term depends solely on the global phase difference, $\Phi$. This term and the bichromatic injection term, which can be zero for equal-amplitude bichromatic seed fields, together contribute to the non-negative values of the second-order moment (SOM), indicating classicality. In contrast, the contributions from spontaneous parametric down-conversion (SPDC) and stimulated parametric down-conversion (StPDC) processes lead to negative SOM values, indicating nonclassicality. 

Our findings reveal the crucial role of $\Phi$: it governs the interplay between classical (through FC and bichromatic injection) and nonclassical (through SPDC and StPDC) contributions, influencing the quantum-to-classical transition in this bipartite system. For instance, when both the phase-mismatching and balanced seeding conditions are met, the FC and bichromatic injection terms become zero simultaneously. This results in strong nonclassicality of the two-mode SSCS across a broad range of squeezing and seeding parameters.

\section{Two-mode Stabilized Squeezed Coherent State}\label{sec:IH}

\begin{figure}[ht]
	\centering
	\includegraphics[width=\columnwidth]{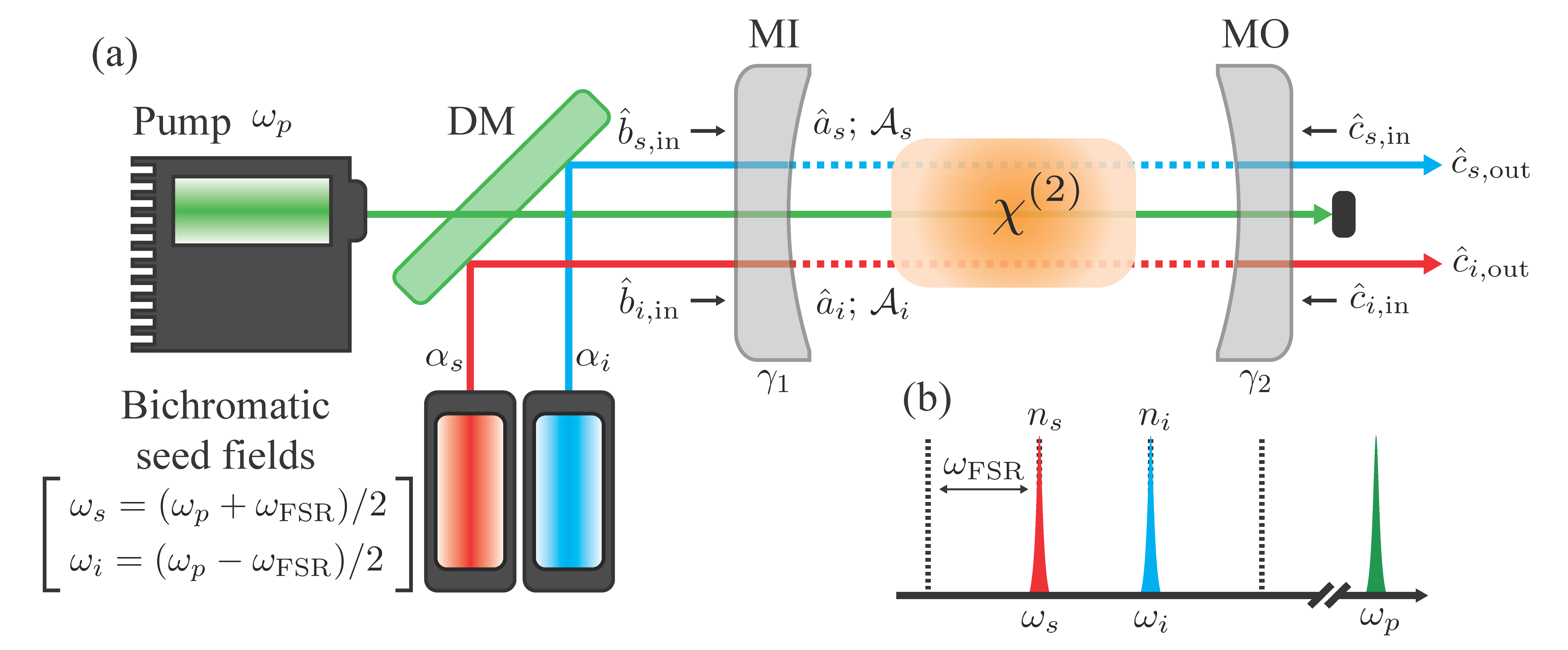}
	\caption{(a) Schematic diagram of a model system to generate a two-mode stabilized squeezed coherent state (SSCS). Various bosonic operators and cavity parameters are defined in the text. (b) Optical spectra of the signal mode (red color, frequency $\omega_s$, mode number $n_s$) and idler mode (blue color, frequency $\omega_i$, mode number $n_i$), separated symmetrically by $\omega_{\rm FSR}$/2 with respect to $\omega_p/2$, where $\omega_p$ (green color) is the pump frequency and $\omega_{\rm FSR}$ is free spectral range of cavity, i.e., their center frequency is exactly given by $\omega_p/2 = (n_s+n_i)\, \omega_{\rm FSR}/2 = (\omega_s+\omega_i)/2$. }\label{Fig0}
\end{figure}

Figure~\ref{Fig0}(a) shows the schematic diagram of a model system to generate a two-mode stabilized squeezed coherent state (SSCS), $|\psi_{\rm SSCS}\rangle$. A nonresonant pump laser (green color, frequency $\omega_p$, phase $\phi_p$ at the input mirror) interacts parametrically with the $\chi^{(2)}$ nonlinear crystal located within the OPO cavity to create a pair of nondegenerate two photons in the signal and idler modes via the SPDC process. We assume that the created signal (dotted red color) and idler (dotted blue color) photons are associated with intra-cavity photon annihilation operators $\hat{a}_s$ and $\hat{a}_i$, respectively, and their frequencies are exactly same as the longitudinal mode frequencies with the same polarization, e.g., using a type-0 phase-matching crystal, at $\omega_s$ and $\omega_i$ as shown in Fig.~\ref{Fig0}(b). The optical spectra of the signal mode (mode number $n_s$) and idler mode (mode number $n_i$) are separated symmetrically by $\omega_{\rm FSR}$/2 with respect to $\omega_p/2$, i.e., their center frequency is exactly given by $\omega_p/2 = (n_s+n_i)\,\omega_{\rm FSR}/2 = (\omega_s+\omega_i)/2$, where the pump frequency for intra-cavity SPDC process is $\omega_p$ (green color) and $\omega_{\rm FSR}$ is free spectral range of cavity.

A phase-matched bichromatic seed field (BSF), i.e., signal and idler seed field with a complex amplitude $\alpha_j = |\alpha_j|e^{i\phi_j}, j\in\{s,i\}$, couples into the OPO cavity through the input mirror (MI) with the same decay rate $\gamma_1$, resulting in the intra-cavity injection parameter for signal and idler modes in the interaction picture as $\mathcal{A}_j =  \sqrt{\gamma_1 |\alpha_j|^2/\Delta t}\,e^{i\phi_j}$, where  $\Delta t $ is the interaction time. In addition, two-mode vacuum fields with input operators $\hat{b}_{s,{\rm in}}$ and $\hat{b}_{i,{\rm in}}$ couple into the two cavity modes through the same MI. Moreover, vacuum fields with input operators $\hat{c}_{s,{\rm in}}$ and $\hat{c}_{i,{\rm in}}$ couple into the OPO cavity through the output mirror (MO) with decay rate $\gamma_2 \gg \gamma_1$. At the same time, the generated two-mode SSCS field output through the same MO with output operators $\hat{c}_{s,{\rm out}}$ and $\hat{c}_{i,{\rm out}}$, respectively. Thus, in our bichromatic seeding method (BSM), we assume that two-mode SSCS is generated by the stimulated PDC process at the pump non-depletion limit, i.e., well below the OPO threshold. The resultant two-mode SSCS field can be easily separated, e.g., by using a mode-filter cavity for bichromatic homodyne detection (see \cite{Marino07,Tian20}).

The two-mode system Hamiltonian in the interaction picture for the system depicted in Fig.~\ref{Fig0}, denoted by $\hat{\mathcal{H}}_{\rm si}$, can be expressed within the two-mode Hilbert space $\mathcal{H}_s\otimes\mathcal{H}_i$ as
\begin{subequations}\label{Hamil}
	\begin{align}
		\hat{\mathcal{H}}_{\rm si}&=\hat{\mathcal{H}}_{\rm DC}+\hat{\mathcal{H}}_{\rm inj}, \label{Hsys}\\
		\hat{\mathcal{H}}_{\rm DC}&=\frac{i\hbar}{2}(\gamma_s^*\hat{a}_s\hat{a}_i-\gamma_s\hat{a}_s^\dagger\hat{a}_i^\dagger),\label{Hdc}\\
		\hat{\mathcal{H}}_{\rm inj}&=i\hbar(\mathcal{A}_s\hat{a}_s^\dagger-\mathcal{A}_s^*\hat{a}_s)+i\hbar(\mathcal{A}_i\hat{a}_i^\dagger-\mathcal{A}_i^*\hat{a}_i),\label{Hin}
	\end{align}
\end{subequations}
where $\hat{\mathcal{H}}_{\rm DC}$ is the two-mode down conversion Hamiltonian with the pump-field coupling (squeezing) parameter $\gamma_s = |\gamma_s|e^{i\phi_p}$ and $\hat{\mathcal{H}}_{\rm inj}$ is the injection Hamiltonian. Here, we assume below threshold condition, i.e., $0<|\gamma_s| < \gamma_{12} = \gamma_1+\gamma_2 \simeq \gamma_2$.

Now, we briefly describe the basic stabilization mechanism for the single-mode system Hamiltonian that has both the squeezing and injection Hamiltonians like Eq.~(\ref{examH}) below \cite{Countinho05,Golubeva08,Puri17,Ruiz18}. 
We consider a single-mode system Hamiltonian $\hat{\mathcal{H}}_{\rm sm}$ as  
	\begin{equation}
		\hat{\mathcal{H}}_{\rm sm} = i\hbar\left( g^*\hat{a}^2-g \hat{a}^{\dagger 2}\right) +i\hbar\left( \mathcal{A} \hat{a}^\dagger - \mathcal{A}^*\hat{a}\right),  \label{examH}
	\end{equation}
where  $g$ is the single-mode squeezing parameter and $\mathcal{A}$ is the injection parameter. Then, under the unitary transformation with $\hat{D}(\alpha) =\exp[\alpha \hat{a}^\dagger - \alpha^* \hat{a}]$, where $\alpha$ is an arbitrary amplitude, the transformed Hamiltonian $\hat{\mathcal{H}}'_{\rm sm} = \hat{D}^\dagger(\alpha) \hat{\mathcal{H}}_{\rm sm}\hat{D}(\alpha)$ can be written as 
	\begin{align}
		\hat{\mathcal{H}}'_{\rm sm} &= i\hbar(g^*\hat{a}^2-g \hat{a}^{\dagger2}) \nonumber \\
		& +\, i\hbar [(2\alpha g^*-\mathcal{A}^*)\hat{a} - (2\alpha^*g-\mathcal{A})\hat{a}^\dagger]  \nonumber \\
		& +\, i\hbar[(g^*\alpha-\mathcal{A}^*)\alpha - (g\alpha^*-\mathcal{A})\alpha^*]. 
	\end{align}
	
To make the transformed Hamiltonian $\hat{\mathcal{H}}'_{\rm sm}$
independent of the first-order bosonic operators in the displaced frame \cite{Puri17}, i.e., representing only the squeezing Hamiltonian with the same squeezing rate $g$, we find that the arbitrary displacement $\alpha$ should take the coherent amplitude 
	\begin{equation}
		\alpha = \frac{\mathcal{A}^*}{2g^*},  \label{newalpha}
	\end{equation} 
and, in this frame, the transformed Hamiltonian has the shifted energy $i\hbar\left({\mathcal{A}^2}/{4g}-{\mathcal{A}^{*2}}/{4g^*}\right)$.
	
As demonstrated above, when the system Hamiltonian contains both the squeezing and injection Hamiltonians like {Eq.~(\ref{examH})}, we can transform it unitarily into an effective system Hamiltonian. To uncouple the system Hamiltonian consisting of the coupled intra-cavity bosonic operators appearing in Eq.~(\ref{Hamil}) into the direct sum of two uncoupled system Hamiltonians, we define two commuting bosonic symmetric annihilation operator  
$\hat{A}_1$ and anti-symmetric annihilation operator $\hat{A}_2$ as  
\begin{equation}
	\hat{A}_1 = \frac{1}{\sqrt{2}} \left(\hat{a}_s+\hat{a}_i\right) \text{ and } \hat{A}_2 = \frac{1}{\sqrt{2}} \left(\hat{a}_s-\hat{a}_i\right).\label{A1A2}
\end{equation}
We see that $[\hat{A}_l,\hat{A}_m^\dagger]=\delta_{lm}, \, \{l,m\}\in\{1,2\}$, where $\delta_{lm}$ is the Kronecker delta and $[\hat{A}_1,\hat{A}_2] = [\hat{A}_1^\dagger,\hat{A}_2^\dagger] = 0$. Next, we define the position $\hat{X}_k$ and momentum $\hat{P}_k$ quadrature operators as  
\begin{equation}
\hat{X}_k=\frac{1}{\sqrt{2}}(\hat{A}_k+\hat{A}_k^\dagger) \text{ and } \hat{P}_k=\frac{i}{\sqrt{2}}(\hat{A}^\dagger_k-\hat{A}_k), \label{QXP}
\end{equation}  
where $[\hat{X}_k,\hat{P}_k]=i$, $k\in\{1,2\}$. Note here that our new bosonic operators are equivalent to the relative position and total momentum quadrature operators in \cite{Fan94}. Similarly, we define two injection rates $\beta_1$ and $\beta_2$ associated with $\mathcal{A}_s$ and $\mathcal{A}_i$ as follows $\beta_1 = \left(\mathcal{A}_s+\mathcal{A}_i\right)/{\sqrt{2}}$ and $\beta_2 = \left(\mathcal{A}_s-\mathcal{A}_i\right)/{\sqrt{2}}$.

Then, the system Hamiltonian $\hat{\mathcal{H}}_{\rm si}$ in Eq.~(\ref{Hamil}) can be expressed in this transformed frame as $\hat{\mathcal{H}}_{12}$. After removing the shifted energy, $\hat{\mathcal{H}}_{12}$ decomposes into two completely decoupled terms, $\hat{\mathcal{H}}_1$ and $\hat{\mathcal{H}}_2$, such that
\begin{subequations}\label{newHs}
\begin{eqnarray}
	\hat{\mathcal{H}}_{12} &=& \hat{\mathcal{H}}_1+\hat{\mathcal{H}}_2, \label{newH} \\
	\hat{\mathcal{H}}_1&=&\hat{D}_1^\dagger(-d_1^*)\hat{\mathcal{H}}_{\rm tm1}\hat{D}_1(-d_1^*), \label{newH1} \\
	\hat{\mathcal{H}}_2&=&\hat{D}_2^\dagger(d_2^*)\hat{\mathcal{H}}_{\rm tm2}\hat{D}_2(d_2^*),\label{newH2}
\end{eqnarray}
\end{subequations}	
where the effective displacement amplitudes are $d_1=2\beta_1/\gamma_s$ and $d_2=2\beta_2/\gamma_s$, respectively, for each Hamiltonian  
$\hat{\mathcal{H}}_1$ and $\hat{\mathcal{H}}_2$. Thus, single-mode squeezing Hamiltonians in Eqs.~(\ref{newH1}) and (\ref{newH2}), those are associated with each phase space for $\{\hat{X}_k,\hat{P}_k\}$, $k \in \{1,2\}$, can be written as 

\begin{subequations} \label{tmHs}
\begin{eqnarray}
	\hat{\mathcal{H}}_{\rm tm1} &=& i\frac{\hbar}{4}\left(\gamma_s^*\hat{A}_1^2-\gamma_s\hat{A}_1^{\dagger2}\right), \label {Htm1}\\
	\hat{\mathcal{H}}_{\rm tm2} &=& -i\frac{\hbar}{4}\left(\gamma_s^*\hat{A}_2^2 - \gamma_s\hat{A}_2^{\dagger2}\right). \label{Htm2}
\end{eqnarray}
\end{subequations}

In what follows, we discuss the evolution of two uncoupled quantum states from the two-mode vacuum state $|0,0\rangle_{1,2} =|0\rangle_1\otimes|0\rangle_2$ by the unitary operators formed by two uncoupled Hamiltonians $\hat{\mathcal{H}}_{\rm tm1}$ and $\hat{\mathcal{H}}_{\rm tm2}$ given in Eq.~(\ref{tmHs}). 
In the Sch\"{o}dinger picture, the two-mode quantum state $|\psi(\tau)\rangle$ after the interaction with the Hamiltonian $\hat{\mathcal{H}}_{12}$ during time $\tau$ can be written as 
\begin{equation}
|\psi_{\rm SSCS}\rangle = \exp\left[-\frac{i\hat{\mathcal{H}}_{12}}{\hbar}\tau\right]|0,0\rangle_{1,2} 
= e^{i \varphi}|\psi_1(\tau)\rangle\otimes|\psi_2(\tau)\rangle, \label{taustate}
\end{equation}
where $\varphi = i [\xi^*(d_1^{*2}-d_2^{*2})-\xi (d_1^2-d_2^2)]/4$. In the last term of Eq.~(\ref{taustate}), $|\psi_1(\tau)\rangle$ and $|\psi_2(\tau)\rangle$ are, respectively,  quantum states associated with the symmetric operator $\hat{A}_1$ and anti-symmetric operator $\hat{A}_2$ and they can be obtained from Eq.~(\ref{newHs}) as 
\begin{subequations}\label{psitau}
\begin{eqnarray}
	|\psi_1(\tau)\rangle&=&\hat{D}_1^\dagger(-d_1^*)\hat{S}_{\rm tm1}(\xi/2)\hat{D}_1(-d_1^*)|0\rangle_1, \label{taus1}\\
	|\psi_2(\tau)\rangle&=&\hat{D}_2^\dagger(d_2^*)\hat{S}_{\rm tm2}(-\xi/2)\hat{D}_2(d_2^*)|0\rangle_2, \label{taus2}
\end{eqnarray} 
\end{subequations}   
where $\xi = \gamma_s \tau$ is the complex squeezing parameter and $\hat{S}_{{\rm tm}k}(\eta)=\exp[\eta^*\hat{A}_k^2/2-\eta \hat{A}_k^{\dagger 2}/2]$, $k\in\{1,2\}$, is the single-mode squeezing operator. Note here that the states in Eq.~(\ref{psitau}) can be interpreted as the unitary-transformed single-mode squeezed states in each phase space associated with the operators $\hat{A}_k$. Indeed, the two states $|\psi_1(\tau)\rangle$ and $|\psi_2(\tau)\rangle$ are obtained, respectively, from the single-mode squeezing Hamiltonians in Eqs.~(\ref{newH1}) and (\ref{newH2}) by applying the unitary operators $\hat{D}_1(d_1)$ and $\hat{D}_2(d_2)$ with effective displacements $d_1$ and $d_2$.

Finally, we want to point out that $|\psi_{\rm SSCS}\rangle$ in Eq.~(\ref{taustate}) is indeed equivalent quantum  state as the displaced squeezed state $|\psi_{\rm DSS}\rangle = \hat{D}(\upsilon_s,\upsilon_i)\hat{S}_{\rm tm}(\xi)|0,0\rangle_{s,i}$, where $\hat{S}_{\rm tm}(\xi) = \exp\left[(\xi^*\hat{a}_s\hat{a}_i -\xi\hat{a}_s^\dagger\hat{a}_i^\dagger)/2  \right]$ is the two-mode squeezing operator \cite{Birrittella15,Birrittella20}. Here, the amplitudes $\upsilon_s$ and $\upsilon_i$ are associated with $d_1$ and $d_2$ in Eq.~(\ref{newHs}) with the Bogoliubov transformation \cite{Lee24}
\begin{subequations}\label{Bog}
\begin{align}
\upsilon_s &= \left(1-\cosh\left(\frac{|\xi|}{2}\right)\right)\mathcal{E}_i^*+e^{i\phi_p}\sinh\left(\frac{|\xi|}{2}\right)\mathcal{E}_s, \\
\upsilon_i &= \left(1-\cosh\left(\frac{|\xi|}{2}\right)\right)\mathcal{E}_s^*+e^{i\phi_p}\sinh\left(\frac{|\xi|}{2}\right)\mathcal{E}_i, 
\end{align}
\end{subequations}
where $\mathcal{E}_{s} = (d_1 + d_2)/\sqrt{2}$ and $\mathcal{E}_{i} = (d_1 - d_2)/\sqrt{2}$. Indeed, with the Bogoliubov transformation in Eq.~(\ref{Bog}), we can show that $|\psi_{\rm SSCS}\rangle  = \hat{D}^\dagger(\mathcal{E}_s,\mathcal{E}_i)\hat{S}_{\rm tm}(\xi)\hat{D}(\mathcal{E}_s,\mathcal{E}_i)|0,0,\rangle_{s,i} = |\psi_{\rm DSS}\rangle$ with the help of Eq.~(\ref{psitau}). Even though the two-mode SSCS in Eq.~(\ref{taustate}) belongs to one of the Gaussian states in the transformed frame with the effective displacements $d_1=2\beta_1/\gamma_s$ and $d_2=2\beta_2/\gamma_s$, respectively, the nonclassicality of the state $|\psi_{\rm SSCS}\rangle$ remains unclear. This is because the system is in a steady state balanced (stabilized) with bichromatic seeding and decaying out of the cavity, which will be discussed in the next section.

\section{Nonclassical Properties \label{sec:TMSS}}  
Two-mode Gaussian state $|\psi_{\rm SSCS}\rangle$ discussed in the previous section has the well-known variance squeezing of one quadrature (anti-squeezing of conjugate quadrature) and non-zero displacement in phase space (see Appendix \ref{sec:QLE}). However, it is unclear whether the quantum state $|\psi_{\rm SSCS}\rangle$ has distinct nonclassical correlations compared to the well-known two-mode squeezed vacuum state and provides sufficient two-mode correlations (entanglement) to be useful for quantum measurement experiments. As such, in this section, we shall first investigate the quantum properties \cite{Cabello13,Shafiee24,Vogel14,Sperling09,Kiesel11,Rahimi13,Tan20,Reid86,Agarwal88,Lee90a,Li00,Marino08,Wasak14,Volovich15} of the two-mode SSCS as functions of controllable system parameters by using the nonclassical measures $\Pi_{\rm A}$ and $\Pi_{\rm L}$ introduced by Agarwal \cite{Agarwal88} and Lee \cite{Lee90a}. We then introduce a novel nonclassical indicator, $\Pi_{\rm N}$, which unveils the contributions of four distinct SOM components. Notably, $\Pi_{\rm N}$ exhibits not only global phase dependence but also analytically separates these four key components of nonclassicality. Understanding these components is critical for characterizing the quantum-to-classical transition in the stabilized bipartite entangled systems.

\subsection{Nonclassical measures $\Pi_{\rm A}$ and $\Pi_{\rm L}$\label{sec:NCM}}

The Gaussian analysis in Appendix~\ref{sec:QLE} does not clearly show the effects of the BSF on the nonclassical properties of the two-mode SSCS. Therefore, we need to further clarify the dependence of quantum properties such as nonclassicality on the coherent injection amplitude $\mathcal{A}_{j}$ for given squeezing parameter $\gamma_s$. Intuitively, however, we may expect that the quantum properties of the two-mode SSCS should exhibit classical field characteristics at higher seed field amplitudes above some critical value. Therefore, we believe that there must be a range of coherent BSF amplitudes, within which the generated two-mode SSCS exhibits clear nonclassical features such as EPR-like quantum correlations.

To address this issue, we first use a nonclassical measure $\Pi_{\rm A}$, defined in Eq.~(\ref{E23}) below, based on the classical inequality, i.e., the CSI \cite{Reid86,Agarwal88}. If the state has a quantum correlation between the signal and idler modes, according to $\Pi_{\rm A}$, the CSI must be violated, i.e., $0 < \Pi_{\rm A} <1 $. In this section, we use the calculation results for the analytical expressions of the self- and cross-correlation functions obtained from the solutions of quantum Langevin equations for the $\hat{a}_s(\Omega)$ and $\hat{a}_i(\Omega)$, where $\Omega$ is the offset frequency from $\omega_j, j\in\{s,i\}$, in the Heisenberg picture (see Appendix \ref{sec:QLE} and \ref{sec:AppB}).

We define $\Pi_{\rm A, in}$ as the nonclassicality measure of intra-cavity fields from the CSI \cite{Agarwal88,Reid86} as
\begin{equation}\label{E23}
	\Pi_{\rm A, in} = \frac{\sqrt{\Gamma_s \Gamma_i}}{|\Gamma_{si}|},
\end{equation}
where $\Gamma_j$, $j\in\{s,i,si\}$, are the SOM functions defined in Eqs.~(\ref{S61}), (\ref{S62}), and (\ref{S63}) and their analytical expressions are given in Eqs.~(\ref{S68}), (\ref{S69}) and (\ref{S72}).
Note here that $\Pi_{\rm A,in}$ takes the value $0 < \Pi_{\rm A} < \infty$, and the two-mode coherent state has the value of $1$. Therefore, any quantum state that has the nonclassical measure between $0 < \Pi_{\rm A} <1$ may be interpreted as a kind of nonclassical state, i.e., a state that has strong two-mode quantum correlations.

\begin{figure*}[t]
\centering
\includegraphics[width=\textwidth]{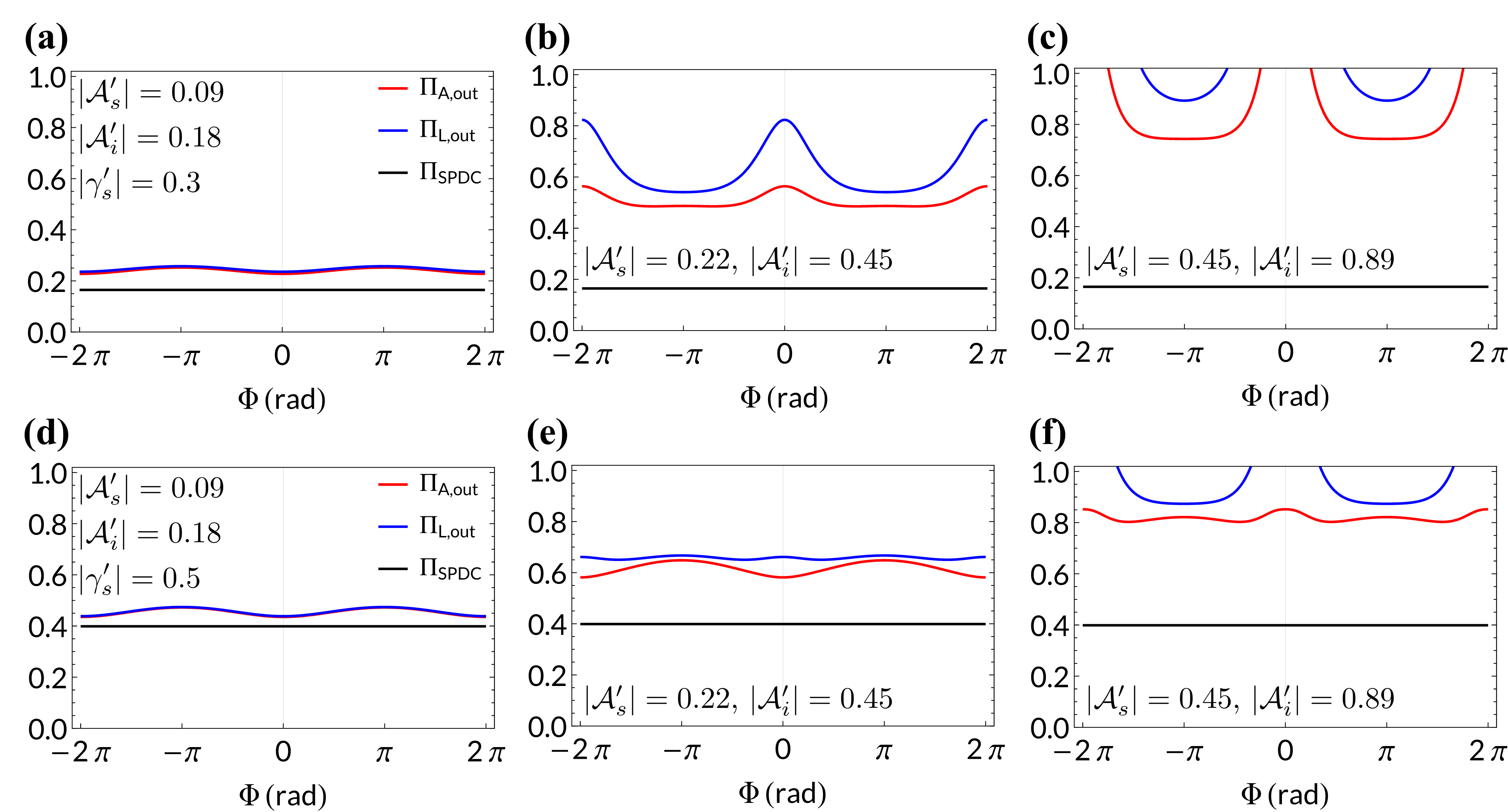}
\caption{$\Pi_{\rm A,out}$ (red solid line) and $\Pi_{\rm L,out}$ (blue solid line) versus $\Phi$ at different seed field amplitudes ($|\mathcal{A}_j'|$). Black solid lines are $\Pi_{\rm SPDC}$ in Eq.~\ref{GSPDC}, which set a lower bound at each squeezing rate. Upper panels (a), (b), and (c) depict $\Pi_{\rm A,out}$ and $\Pi_{\rm L,out}$ for a weak squeezing rate of $|\gamma_s'| = 0.3$, while lower panels (d), (e), and (f) depict them for an intermediate squeezing rate of $|\gamma_s'| = 0.5$. The left column (panel (a) and (d)), middle column (panel (b) and (e)), and right column (panel (c) and (f)) correspond, respectively, to weak, intermediate, and strong seed field amplitudes (provided by $|\mathcal{A}_s'|$ and $|\mathcal{A}_i'|$ values). We used the relation $\gamma_1/\gamma_2 = 0.002$ for the mirror decay rates.
	}\label{FigNCM}
\end{figure*}

For comprehensive discussions, we also consider a similar nonclassical measure $\Pi_{\rm L}$, defined in Eq.~(\ref{EM}) below, based on the Murihead inequality \cite{Lee90a} as
\begin{equation}\label{EM}
	\Pi_{\rm L,in}=\frac{1}{2} \frac{\Gamma_s +\Gamma_i}{|\Gamma_{si}|}.
\end{equation}
The nonclassical measure $\Pi_{\rm L,in}$ also takes the value $0 < \Pi_{\rm L}<\infty$, and any nonclassical state ought to have the value between $0 < \Pi_{\rm L}<1$. Note here that if injection parameters are symmetric, i.e., $\mathcal{A}_s=\mathcal{A}_i$, $\Pi_{\rm L,in}$ is equivalent to $\Pi_{\rm A,in}$.

Furthermore, by using the input-output relation, we can obtain the nonclassical measures $\Pi_{\rm A,out}$ and $\Pi_{\rm L,out}$ of the two-mode SSCS at the output of the OPO cavity. After a lengthy calculation as described in Appendix \ref{sec:AppB}, we find the nonclassical measures of the output fields as
\begin{align}
	\Pi_{\rm A, out}&=\frac{\sqrt{\Gamma_{s}^{\rm out}\Gamma_i^{\rm out}}}{|\Gamma_{si}^{\rm {out}}|} = \Pi_{\rm A, in}, \label{E28} \\
	\Pi_{\rm L, out}&=\frac{1}{2}\frac{\Gamma_{s}^{\rm out}+\Gamma_i^{\rm out}}{|\Gamma_{si}^{\rm {out}}|} = \Pi_{\rm L, in}, \label{E28b}
\end{align}
where the individual two-mode correlation functions are found to be
	\begin{equation} \label{E29}
		\Gamma_s^{\rm out} =\gamma_2^2 \Gamma_s,  
		\Gamma_i^{\rm out} =\gamma_2^2\Gamma_i,  {\text{ and }}
		\Gamma_{si}^{\rm out} =\gamma_2^2\Gamma_{si}.   
	\end{equation}

The surprising equivalences of $\Gamma_s^{\rm out}$ and $\Gamma_i^{\rm out}$
to their intra-cavity counterparts, found in Eqs.~(\ref{E28}) and (\ref{E28b}), can be understood as follows. Since $\Gamma_s^{\rm out}$ and $\Gamma_i^{\rm out}$
all relate to the intra-cavity ones through the input-output relations in Eq.~(\ref{E15}), they share a common multiplicative factor of $\gamma_{2}^2$ due to passing through the same output mirror MO with decay rate $\gamma_{2}$. Consequently, the ratio in Eq.~(\ref{E28}) cancels the effects of the output mirror coupling rate $\gamma_{2}$, leading to the identical nonclassical measure:  $\Pi_{\rm A, out} = \Pi_{\rm A, in}$ and $\Pi_{\rm L, out} = \Pi_{\rm L, in}$. In other words, the OPO cavity does not enhance or destroy the SOM property of the intra-cavity fields.

On the other hand, we note that the intra-cavity squeezing level has the maximum value of -3 dB corresponding to the squeezing parameter $r_{\rm int}=-\frac{1}{2}\ln \left[1-|\gamma_s'|/(1+|\gamma_s'|)\right] = 0.35$ for $|\gamma_s'| = 1$ at OPO threshold \cite{Collett84,Qin22}, where $|\gamma_s'| = |\gamma_s|/\gamma_{12}$. As contrary to the nonclassical measure, however, the extra-cavity squeezing parameter $r_{\rm ext}$, which can be enhanced by the OPO cavity, may have the value up to the infinite squeezing \cite{Collett84,Qin22}, and it is given as 
\begin{equation}
r_{\rm ext} = -\frac{1}{2}\ln\left[\frac{V_{\rm out}(0)}{V_{\rm 0}}\right] =- \frac{1}{2}\ln \left[1-\frac{4|\gamma_s'|}{(1+|\gamma_s'|)^2}\right],
\end{equation}
where $V_{\rm out}(\Omega)$ is the two-mode output variance given in Eq.~(\ref{E19}) and $V_{\rm 0}=1/2$ is the two-mode vacuum  variance.

Figure~\ref{FigNCM} depicts $\Pi_{\rm A,out}$ and $\Pi_{\rm L,out}$ as a function of the global phase difference $\Phi = \phi_s+\phi_i - \phi_p$ for various BSF amplitudes $|\mathcal{A}_j'|$ at two different squeezing rates: $|\gamma_s'| = 0.3$ (low squeezing rate in (a), (b), and (c)) and 0.5 (high squeezing rate in (d), (e), and (f)) with asymmetric injection rates $|\mathcal{A}_i'|=2|\mathcal{A}_s'|$. Here, we used dimensionless parameters such as $|\mathcal{A}_j'|=|\mathcal{A}_j|/\gamma_{12}, j\in\{s,i\}$. 

We note here that $\Pi_{\rm A,out}$ and $\Pi_{\rm L,out}$ 
for the SPDC process (denoted by $\Pi_{\rm SPDC}$) do not depend on $\Phi$, since in that case there is no bichromatic seed fields, i.e., $\mathcal{A}_s = \mathcal{A}_i = 0$ in $\Gamma_s, \Gamma_i$, and $\Gamma_{si}$.  Thus, $\Pi_{\rm SPDC}$ has the minimum value of $\Pi_{\rm A,out}$ and $\Pi_{\rm L,out}$ at the given squeezing parameter $|\gamma_s'|$ (black solid lines in Fig.~\ref{FigNCM}), and has a very simple expression as
\begin{equation}
	\Pi_{\rm SPDC}=\frac{2|\gamma_s'|^2}{1+|\gamma_s'|^2}. \label{GSPDC}
\end{equation}
In our BSM presented in Fig.~\ref{Fig0}, the squeezing rate takes on values in the range $0 <|\gamma_s'| < 1$. This indicates that our linear model in Eq.~(\ref{E1E2}) operates below the OPO threshold (where $|\gamma_s'| = 1$). Consequently, the range of the nonclassical measure $\Pi_{\rm SPDC}$ falls within 0 and 1. Interestingly, $\Pi_{\rm SPDC}$ increases monotonically as $|\gamma_s'|$ increases, similar to the observed behavior in the macroscopic coherent regime \cite{Marino08}.

In weak seeding regimes in Fig.~\ref{FigNCM} with $|\mathcal{A}_s'|=0.09$ and $|\mathcal{A}_i'|= 2|\mathcal{A}_s'|= 0.18$, $\Pi_{\rm A,out}$ and $\Pi_{\rm L, out}$ exhibit almost the same weak dependence on $\Phi$, reaching their minimum value (maximum nonclassicality) at $\Phi=2 n\pi, \, n\in \mathbb{Z}$, for $|\gamma_s'|$ = $0.3$ in panel (a) and 0.5 in panel (d). However, when we increase the seed beam amplitudes to $|\mathcal{A}_s'|=0.22$ (panel (b) and (e)) and $0.45$ (panel (c) and (f)), $\Pi_{\rm A, out}$ and $\Pi_{\rm L,out}$ deviate from each other more significantly, and the global phase $\Phi$ at which the nonclassical measures have their minimum values changes from $\Phi=2n\pi$ to $(2n+1)\pi$. In these regimes, the system undergo a quantum-to-classical transition much faster at $\Phi = 2n\pi$ compared to the one at $\Phi = (2n+1)\pi$. This behavior can be understood by considering that perfect phase matching of the PDC occurs at $\Phi = 2n\pi$, so the coherent seed field is amplified rapidly by the classical frequency conversion (FC) process and reaches the classical boundary easier.

Furthermore, in the intermediate seeding regime (panels (b) and (e)), the nonclassical measures have values still less than 1 for whole range of $\Phi$, confirming that the system exhibits nonclassical quantum correlations at any $\Phi$. However, in the strong seeding regime (panels (c) and (f)), the nonclassical measures exceed 1 at $\Phi=2n\pi$, suggesting that the system has undergone a quantum-to-classical transition at these specific phase values. We also observe that the lower bound of the nonclassical measure approaches 1 (classical) as the squeezing rate $|\gamma_s|$ approaches the OPO threshold value, $\gamma_{12}$. It is interesting to note from our analytical expressions in Eqs.~(\ref{S68}), (\ref{S69}) and (\ref{S72}) that $\Pi_{\rm A, out}$ and $\Pi_{\rm L,out}$ are symmetric with respect to the amplitude ratio of the bichromatic seed fields. Moreover, across all parameter ranges depicted in Fig.~\ref{FigNCM}, $\Pi_{\rm L,out}$ consistently exceeds $\Pi_{\rm A,out}$ for whole range  of $\Phi$. This suggests that $\Pi_{\rm L,out}$ can serve as a more sensitive indicator of the quantum-to-classical transition compared to $\Pi_{\rm A,out}$.

\begin{figure}[ht]
	\centering
	\includegraphics[width=\columnwidth]{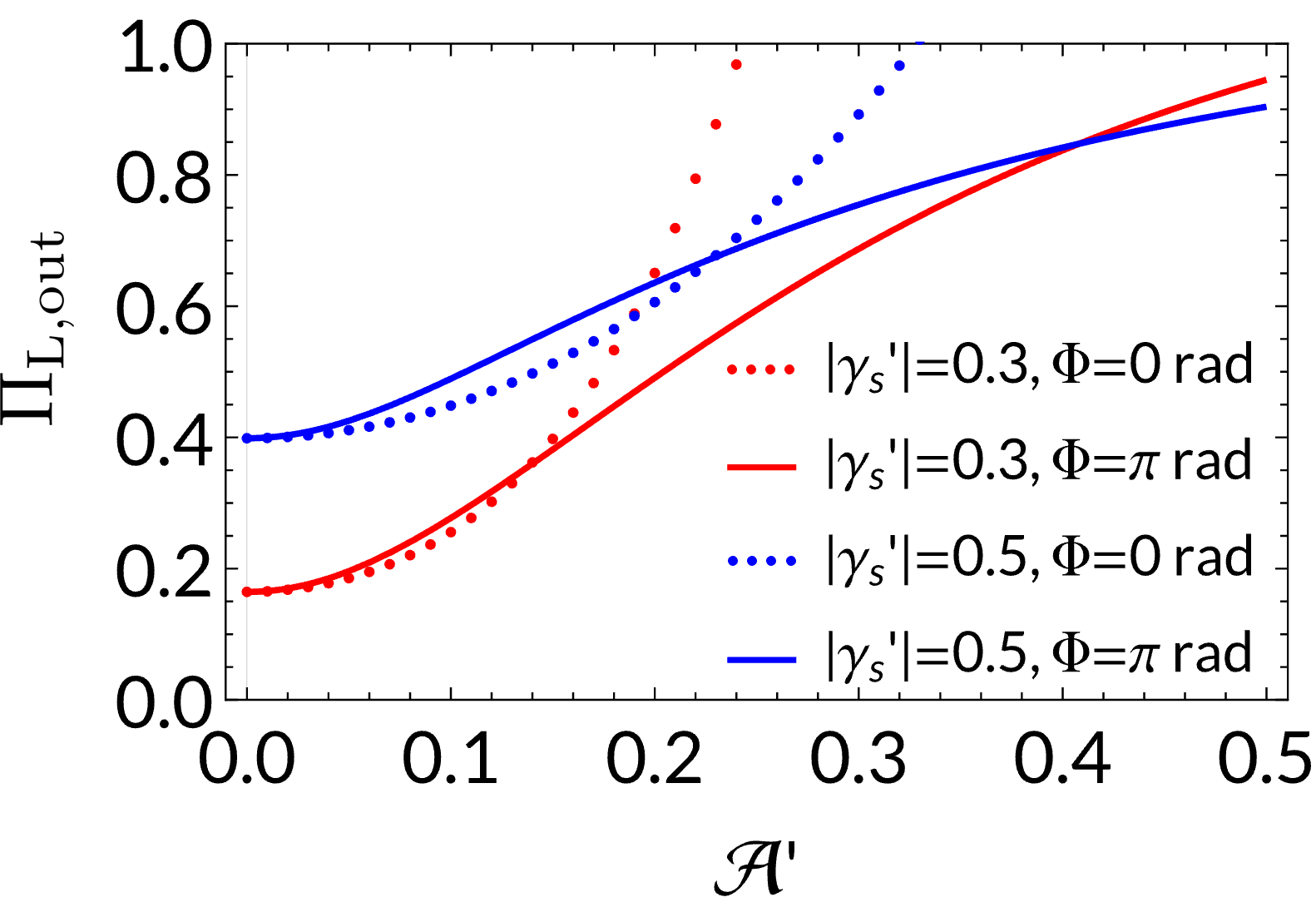}
	\caption{
		$\Pi_{\rm L,out}$ versus $\mathcal{A'} = \sqrt{|\mathcal{A}_i'|^2+|\mathcal{A}_s'|^2}$
		with $|\mathcal{A}_i'| = 0.1$ at $|\gamma_s'|=0.3$ and $0.5$ (dotted and dashed lines for $\Phi = 0$, and dash-dotted and solid lines for $\Phi = \pi$ rad, respectively) with$\gamma_1/\gamma_2=0.002$.}\label{FigNCM2}
\end{figure}

In Fig.~\ref{FigNCM2}, we further analyze the dependence of the seed amplitude on $\Pi_{\rm L,out}$ at two particular values of $\Phi = 0$ and $\pi$ rad plotted against the total injection rate $\mathcal{A'} = \sqrt{|\mathcal{A}_i'|^2+|\mathcal{A}_s'|^2}$ for two different  $|\gamma_s'|$: $|\gamma_s'| = 0.3$ (red dotted line for $\Phi = 0$ and red solid line for $\Phi = \pi$ rad) and $|\gamma_s'| = 0.5$ (blue dotted line for $\Phi = 0$ and blue solid line for $\Phi = \pi$ rad). Note here that at $\mathcal{A'} = 0 $ $\Pi_{\rm L, out} = \Pi_{\rm SPDC}  \simeq 0.16$ and 0.40 for $|\gamma_s'| = 0.3$ and 0.5, respectively. At perfect phase-matching condition for the PDC process in a $\chi^{(2)}$ nonlinear crystal, i.e.,  $\Phi = 0$, $\Pi_{\rm L,out}$ increases across 1 at $\mathcal{A'} = 0.24$ for $|\gamma_s'| = 0.3$ and $\mathcal{A'} = 0.32$ for $|\gamma_s'| = 0.5$ as $\mathcal{A'}$ increases from 0. At $\Phi = \pi$, however, $\Pi_{\rm L,out}$ remains less than 1 as $\mathcal{A'}$ increases up to 0.5 for both values of $|\gamma_s'|$, confirming the observations in Fig.~\ref{FigNCM} that at $\Phi = 0$, $\Pi_{\rm L, out}$ makes a quantum-to-classical transition faster than at $\Phi = \pi$ rad as $\mathcal{A'}$ increases from 0. However, it is unclear which process (or system parameter) dictates this quantum-to-classical transition most strongly when controllable system parameters of $|\gamma_s'|$ and $|\mathcal{A}_j'|, j \in \{s,i\}$ are varied.

\subsection{Novel nonclassicality indicator $\Pi_{\rm N}$\label{sec:PiN}}

To investigate the effects of each physical process on the nonclassical measure in our BSM, we introduce a novel nonclassical indicator $\Pi$ in Eq.~(\ref{Lambda}) below. This measure is derived from the difference between the arithmetic mean and the geometric mean in Eq.~(\ref{EM}) as    
\begin{align}
	\Pi=\frac{1}{2}\left(\Gamma_s+\Gamma_i\right)-\Gamma_{si}. \label{Lambda}
\end{align}

Then, $\Pi$ differs fundamentally from $\Pi_{\rm L}$. Unlike $\Pi_{\rm L}$
which is defined by the ratio of the arithmetic and geometric means,  $\Pi$ is defined by their difference. This allows  $\Pi$ to have values outside the 0 to 1 range. A negative value of  $\Pi$ (i.e., $-\infty < \Pi < 0$) indicates nonclassicality of the two-mode state. In this region, the cross-correlation function $\Gamma_{si}$ has a larger value than the self-correlation functions $\Gamma_s$ and $\Gamma_i$. In other words, the quantum-to-classical transition occurs at the point where $\Pi = 0$. When the system exhibits predominantly nonclassical properties, $\Pi$ will be negative (less than zero). Conversely, for a mostly classical system, $\Pi$ will be non-negative (greater than or equal to zero). 

In particular, $\Pi$ can be decomposed into four individual terms, where each term captures the influence of a specific physical process. Utilizing the notation ${\rm Qs'}^2 = 4/(1-|\gamma_s'|^2)^2$, we can express $\Pi$ as
\begin{equation}\label{PiN}
\Pi={\rm Qs'}^2\left(\Pi_{\rm inj}+\Pi_{\rm FC}+\Pi_{\rm SPDC}+\Pi_{\rm StPDC}\right),
\end{equation}
where $\Pi_{\rm inj}$, $\Pi_{\rm FC}$, $\Pi_{\rm SPDC}$, and $\Pi_{\rm StPDC}$ are terms corresponding to each physical process occurring in our BSM in Fig.~\ref{Fig0} such as BSF injection, nonlinear FC, SPDC, and StPDC, respectively. In Eq.~(\ref{PiN}), each term has explicit analytical expressions as
\begin{subequations}\label{PiY}
\begin{align}
	&\Pi_{\rm inj}=8\pi^2  \left(|\mathcal{A}_s'|^2-|\mathcal{A}_i'|^2\right)^2,\label{inj}\\
	&\Pi_{\rm FC}=4\pi^2  |\gamma_s'|\,|\mathcal{A}_s'|\,|\mathcal{A}_i'|\left(1+\cos\Phi\right),\label{FC}\\
	&\Pi_{\rm SPDC} = -\frac{\pi^2}{4}|\gamma_s'|^2\left(1-|\gamma_s'|^2\right),\label{SPDC}\\
	&\Pi_{\rm StPDC}=-2\pi^2|\gamma_s'|^2\left(|\mathcal{A}_s'|^2+|\mathcal{A}_i'|^2\right) -4\pi^2  |\gamma_s'|\,|\mathcal{A}_s'|\,|\mathcal{A}_i'|.\label{StPDC}
\end{align}
\end{subequations}

Each term in Eq.~(\ref{PiY}) has its own physical origin. The first term, $\Pi_{\rm inj} \ge 0$ always (classicality measure), depends only on the difference between $|\mathcal{A}_s'|$ and $|\mathcal{A}_i'|$, i.e., it depends critically on the asymmetry of seeding amplitudes. If $|\mathcal{A}_s'| = |\mathcal{A}_i'|$, there is no contribution to $\Pi$, meaning $\Pi_{\rm inj}$ has always non-negative contribution to $\Pi$. The second term, $\Pi_{\rm FC} \ge 0$ always (classicality measure), is proportional to $|\gamma_s'\mathcal{A}_s'\mathcal{A}_i'|(1+\cos\Phi)$, representing the coherent frequency conversion process with phase-matching factor of $\cos\Phi$. At $\Phi = \pi$ rad, $\Pi_{\rm FC} = 0$, thus $\Pi_{\rm FC}$ has always non-negative contribution to $\Pi$. We can now understand the $\Phi$ dependence of $\Pi_{\rm A}$ and $\Pi_{\rm L}$ in Fig.~\ref{FigNCM}, confirming that at $\Phi = \pi$ rad, they have more nonclassicality than the case when $\Phi = 0$, and thus the quantum-to-classical transition occurs first at $\Phi = 0$.  

The third term, $\Pi_{\rm SPDC} < 0$ always for $0 < |\gamma_s'| < 1$ (nonclassicality measure), depends only on $|\gamma_s'|^2$, indicating that the SPDC process occurs with the parametric pump field alone. Finally, the last term, $\Pi_{\rm StPDC} < 0$ always (nonclassicality measure) if $|\mathcal{A}_j'| \ne 0$. This term describes a stimulated single-photon and photon-pair emission processes with the pump and the BSF similar to $\Pi_{\rm FC}$, but it does not depend on $\Phi$ due to the nature of the StPDC process. We want to emphasize that this capacity of revealing contributions of individual physical processes in the nonclassicality measure distinguishes our novel indicator $\Pi$ from any previous nonclassical measures.

Many features can be drawn from Eq.~(\ref{PiN}), since we know analytically the contributions of each term to $\Pi$. Most importantly, the global phase dependence observed in Fig.~\ref{FigNCM} is due to the contribution of $\Pi_{\rm FC}$, because it is the only term that depends on $\Phi$. Moreover, $\Pi_{\rm inj}$ is always positive, making the state more classical as the square of the BSF intensity increases. On the other hand, $\Pi_{\rm SPDC}$ is always negative, and it sets the minimum value of $\Pi$ (maximum nonclassicality) when the system operates without the BSF in the below-threshold regime, i.e., $|\gamma_s'| < 1$, as same as in our BSM. Similarly, $\Pi_{\rm StPDC}$ always has a negative value and decreases depending on the intensity of the bichromatic seed fields at a given squeezing rate $|\gamma_s'|$. Lastly, all the contributions from the four distinct physical processes sum up to $\Pi$ in Eq.~(\ref{PiN}), which measures the nonclassicality ($\Pi < 0$) of the system with high sensitivity.

As explained above, $\Pi_{\rm SPDC}$ always has a negative value and is a measure of the nonclassicality of the state without the bichromatic seed fields. In particular, it reaches its minimum value at $|\gamma_s'|=1/\sqrt{2}$. Therefore, it is natural to define a normalized nonclassicality indicator, $\Pi_{\rm N}$, by dividing $\Pi$ in Eq.~(\ref{PiN}) by ${\rm Qs'}^2|\Pi_{\rm SPDC}|$ as
\begin{eqnarray}\label{NomPi}
	\Pi_{\rm N}  
		&=& \Pi_{\rm N,inj}+\Pi_{\rm N,FC}+\Pi_{\rm N,SPDC}+\Pi_{\rm N,StPDC} \nonumber \\
		&=& \frac{32\mathcal{A'}^4\cos^2(2\theta)+8\mathcal{A'}^2|\gamma_s'|\left(\sin(2\theta)\cos\Phi-|\gamma_s'|\right)}{|\gamma_s'|^2(1-|\gamma_s'|^2)} - 1, \nonumber \\
\end{eqnarray}
where $\mathcal{A'} = |\mathcal{A}_i'|/\cos\theta$, $\tan\theta = |\mathcal{A}_s'|/|\mathcal{A}_i'|$.  If we consider the condition of maximum nonclassicality of SPDC process, i.e., $|\gamma_s'|=1/\sqrt{2}$, Eq.~(\ref{NomPi}) simply reads
\begin{equation}\label{SimPi}
	\Pi_{\rm N} = 128\mathcal{A'}^4 \cos^2(2\theta)
	+16\mathcal{A'}^2\left(\sqrt{2}\sin(2\theta)\cos\Phi-1 \right)-1.
\end{equation}

\begin{figure}[t]
	\centering
	\includegraphics[width=\columnwidth]{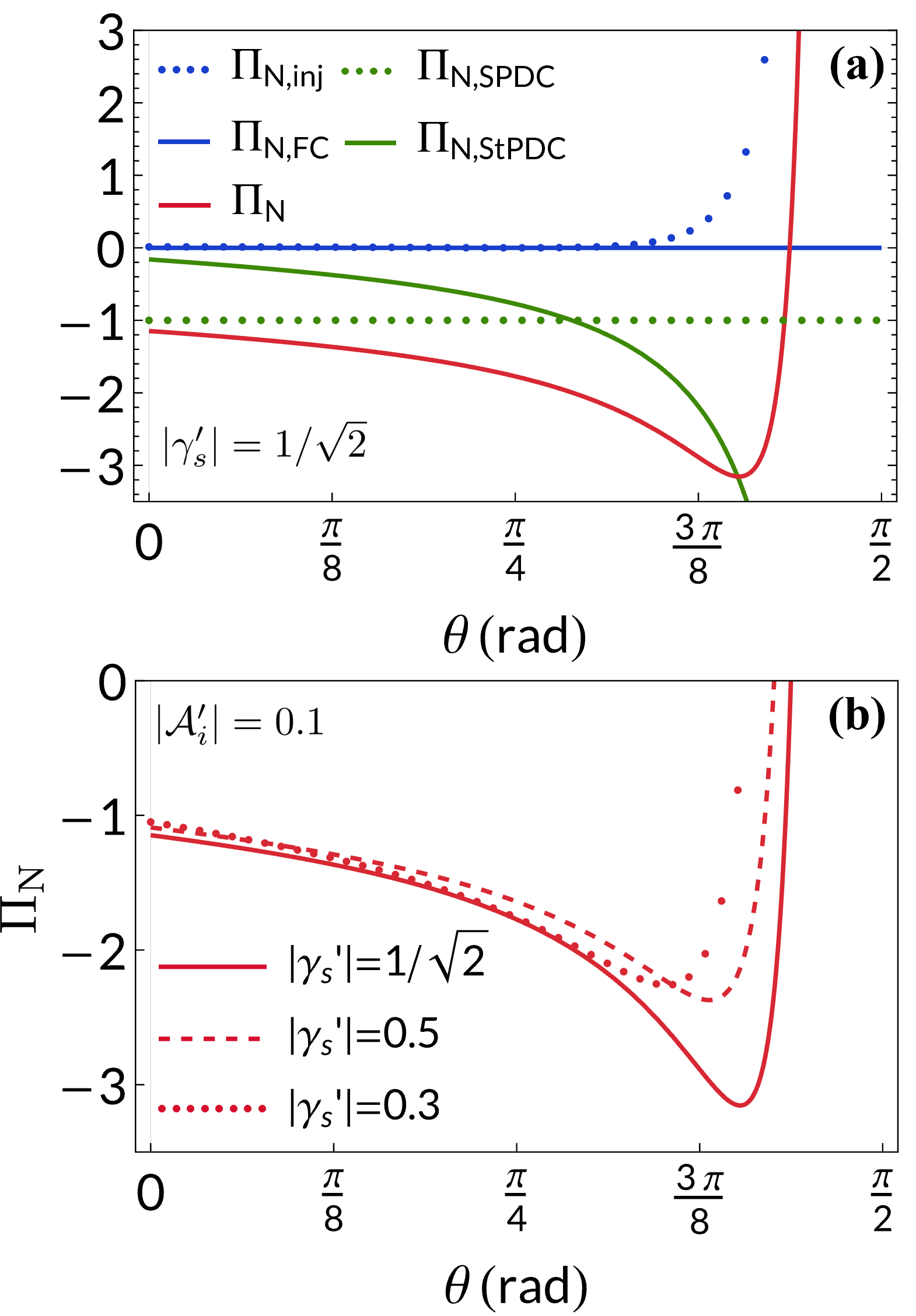}
	\caption{(a) Component $\Pi_{\rm N,inj}$ (blue dotted line), $\Pi_{\rm N,FC}$ (blue solid line), $\Pi_{\rm N,SPDC}$ (green dotted line), $\Pi_{\rm N,StPDC}$  (green solid line), and $\Pi_{\rm N}$ (red solid line) versus $\theta =\tan^{-1}\left(|\mathcal{A}_s|/|\mathcal{A}_i|\right)$ for $|\gamma_s'| = 1/\sqrt{2}$ and $|\mathcal{A}_i'| = 0.1$. (b) $\Pi_{\rm N}$ versus $\theta$ for three different values of the squeezing rate for $|\mathcal{A}_i'| = 0.1$: $|\gamma_s'| = 1/\sqrt{2}$ (red solid line), 0.5 (red dashed line), and 0.3 (red dotted line). 
	}\label{FigNormal}
\end{figure}
Figure~\ref{FigNormal}(a) depicts the dependence of individual components of $\Pi_{\rm N}$ in Eq.~(\ref{NomPi}) at the phase-mismatch condition, i.e., $\Phi = \pi$, on the asymmetric bichromatic seed rate $\theta  = \tan^{-1}\left(|\mathcal{A}_s'|/|\mathcal{A}_i'|\right)$ for $|\mathcal{A}_i'| = 0.1$ and $|\gamma_s'| = 1/\sqrt{2}$: $\Pi_{\rm, N,inj}$ (blue dotted line), $\Pi_{\rm, N,FC}$ (blue solid line), $\Pi_{\rm, N,SPDC}$ (green dotted line), and $\Pi_{\rm, N,StPDC}$ (green solid line). $\Pi_{\rm N}$ (red solid line) is the sum of the four components. At this phase-mismatch condition of $\Phi = \pi$, $\Pi_{\rm N,FC} = 0$ (see Eq.~(\ref{FC})). Moreover, $\Pi_{\rm N,inj}$ in Eq.~(\ref{inj}) always has a positive value, while $\Pi_{\rm N,SPDC}$ and $\Pi_{\rm N,StPDC}$ in Eqs.~(\ref{SPDC}) and (\ref{StPDC}), respectively, have always negative values regardless of the values of squeezing rate $|\gamma_s'|$ and bichromatic seed rates $|\mathcal{A}_j'|$. $\Pi_{\rm N}$ (red solid line) sums up the four contributions from the four physical processes and it has a minimum negative value (maximum nonclassicality) at $\theta = 0.41 \pi$ rad, since $\Pi_{\rm N,inj}$ dominates $\Pi_{\rm N,StPDC}$ for $\theta \ge 0.41 \pi$.

Figure~\ref{FigNormal}(b) shows  $\Pi_{\rm N}$ as a function of $\theta$ for three different squeezing rates: $|\gamma_s'| = 1/\sqrt{2}$ (green solid line), 0.5 (green dashed line), and 0.3 (green dotted line), all for a fixed value of $|\mathcal{A}_i'| = 0.1$. At the maximum nonclassicality condition of $\Pi_{\rm N,SPDC}$, i.e., $|\gamma_s'| = 1/\sqrt{2}$, $\Pi_{\rm N}$ clearly shows the lowest nonclassicality. Specifically, in this high seeding regime ($\theta > 0.41\pi$ rad), $\Pi_{\rm N,inj}$, which is proportional to $|\mathcal{A}_j'|^4$ as can be seen in Eq.~(\ref{inj}), dominates the other three terms,for three other terms ($\Pi_{\rm N,FC}, \Pi_{\rm N,SPDC}$, and $\Pi_{\rm N,StPDC}$ in Eq.~(\ref{PiY})) depend mostly on $|\mathcal{A}_j'|^2$. This dominance by $\Pi_{\rm N,inj}$ causes $\Pi_{\rm N}$ of the quantum state to make a quantum-to-classical transition across a value of zero. 

\begin{figure}[t]
\centering
\includegraphics[width=\columnwidth]{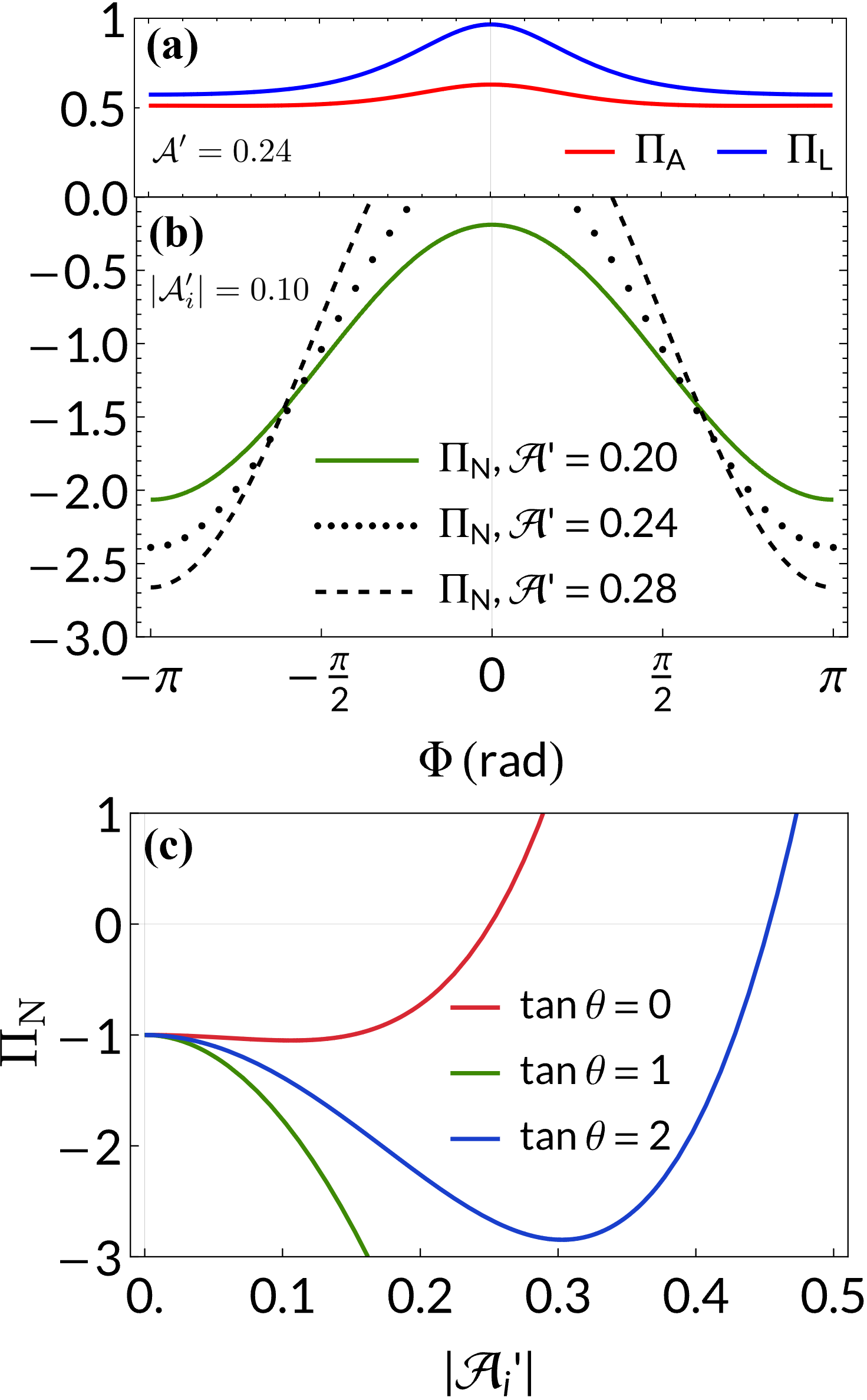}
\caption{(a) $\Pi_{\rm A}$ (red line) and $\Pi_{\rm L}$ (blue line) versus  $\Phi$ at $\mathcal{A'} = 0.24$ and $|\mathcal{A}_i'| = 0.1$ for comparison. 
(b) $\Pi_{\rm N}$ versus $\Phi$ for three different $\mathcal{A}'$: $\mathcal{A'} = 0.20$ (green solid line), 0.24 (black dotted line), and 0.28 (black dashed line).
(c) $\Pi_{\rm N}$ versus $\mathcal{A}_i'$ for three different $\theta$: $\tan\theta = 0$ (red solid line), $\tan\theta = 1$ (green solid line), and $\tan\theta = 2$ (green solid line). In Fig.~\ref{FigCom}, we set $|\gamma_s'| = 0.3$.
	}\label{FigCom}
\end{figure} 

Figure~\ref{FigCom} summarizes the parameter dependence of $\Pi_{\rm N}$. To compare the sensitivity of $\Pi_{\rm N}$ on $\Phi$ with those of $\Pi_{\rm A}$ and $\Pi_{\rm L}$, we show $\Pi_{\rm A}$ and $\Pi_{\rm L}$ versus $\Phi$ in Fig.~\ref{FigCom}(a) at $\mathcal{A} = 0.24$ and $|\mathcal{A}_i'| = 0.1$. In Fig.~\ref{FigCom}(b), we depict the dependence of $\Pi_{\rm N}$ on $\Phi$ for three different values of $\mathcal{A'}$: $\mathcal{A'} = 0.20$ (green solid line), 0.24 (black dotted line), and 0.28 (black dashed line). In all three cases, we fixed the idler seed rate to $|\mathcal{A}_i'| = 0.1$. In Fig.~\ref{FigCom}(c), we compare the dependence of $\Pi_{\rm N}$ on $|\mathcal{A}_i'|$ for three different asymmetric seed ratios $\tan\theta$: $\tan \theta = 0$ (red solid line), i.e., idler seed only, $\tan\theta = 1$ (symmetric seeding, green solid line), and $\tan\theta = 2$ (asymmetric seeding, blue solid line). 

When we increase $\mathcal{A'}$ from 0.20 to 0.24 and 0.28 (while fixing the other parameters), $\Pi_{\rm N}$ shows very similar functional dependence on $\Phi$ as those of $\Pi_{\rm A}$ and $\Pi_{\rm L}$ as shown in panels (a) and (b). At $\Phi = 0$, they are closer to the border of the quantum-to-classical transition. However, at $\Phi = \pi$, they have their minimum possible value due to the effect of $\Pi_{\rm N,FC}$ as discussed in Fig.~\ref{FigNormal}. For example at $\mathcal{A'} = 0.20$ (green solid line), $\Pi_{\rm N}$ still exhibits the nonclassicality of the quantum state for all values of $\Phi$. However, at $\mathcal{A'} = 0.24$, $\Pi_{\rm N} > 0$ at $\pi = 0$, while $\Pi_{\rm A}$ and $\Pi_{\rm L}$ remain in the nonclassical region. This result indicates that our novel measure $\Pi_{\rm N}$ is more sensitive for determining the nonclassicality of EPR-entangled two-mode squeezed light.

In panel (c), the parameters are fixed at $\Phi = \pi$ and $|\gamma_s'| = 0.3$. We see that even in the case of a single seed beam (red solid line for $\tan\theta = 0$)), $\Pi_{\rm N}$ exhibits a broad nonclassical region for $0 < |\mathcal{A}_j'| \simeq 0.25$, which supports the previously reported observation of single-photon interference in Refs.~\cite{Lee18,Yoon21}. Moreover, we also observe that for asymmetric BSF amplitudes,$\Pi_{\rm N}$ achieves its maximum nonclassicality possible for these fixed system parameters, for example at  $|\mathcal{A}_i'| \simeq 0.3$ (blue solid line). We note that one can achieve $\Pi_{\rm N} \ll 0$ for the given system parameters in our BSM of Fig.~\ref{Fig0} by carefully choosing $\Phi = \pi$ and perfectly symmetric seeding (green solid line). In this case, however, even slight fluctuations of $|\mathcal{A}_j'|$ can force the system to experience quantum-to-classical transition due to the dependence of $\Pi_{\rm N,inj}$ on $|\mathcal{A}_j'|^4$.

\section{Conclusions and Outlooks}
This work investigated nonclassicality criteria for a two-mode stabilized squeezed coherent state (SSCS) of light exhibiting high nonclassicality. The SSCS was generated via an OPO cavity driven by bichromatic weak seed fields. We focused on a quantum system comprising the intra- and extra-cavity fields, operating far below the OPO threshold.

We analyzed first established nonclassicality measures,  $\Pi_{\rm A}$ and $\Pi_{\rm L}$, based on the Cauchy-Schwarz and Muirhead inequalities, respectively. Surprisingly, the OPO cavity did not enhance these measures, suggesting that intra-cavity processing does not necessarily improve nonclassicality. Additionally, we found that the global phase difference, $\Phi = \phi_s+\phi_i-\phi_p$, significantly affects these measures. However, for symmetric seeding amplitudes, both $\Pi_{\rm A}$ and $\Pi_{\rm L}$
provided equivalent results. For asymmetric seeding, $\Pi_{\rm L}$ emerged as the more suitable measure.

We then introduced a novel nonclassicality indicator, $\Pi_{\rm N}$. This indicator explicitly incorporates four individual terms representing key physical processes: coherent bichromatic injection, nonlinear FC, SPDC, and StPDC. Notably, $\Pi_{\rm N}$ not only identified the optimal $\Phi$ for maximum nonclassicality, aligning with the existing criteria, but also pinpointed the responsible mechanism – the $\Phi$-dependent nonlinear FC process. This highlights $\Pi_{\rm N}$'s ability to reveal previously hidden aspects of nonclassicality.

Our analysis of the parameter dependence of these four components paves the way for studying the quantum-to-classical transition in two-mode entangled systems. Based on the analytical expressions for each component of $\Pi_{\rm N}$ and numerical simulations, we propose strategies to achieve maximum nonclassicality of the SSCS in the bichromatic seeding method. These strategies involve careful control of the global phase difference ($\Phi$), squeezing rate ($\gamma_s$), and bichromatic seed amplitudes ($\mathcal{A}_s$ and $\mathcal{A}_i$) within a weakly-seeded OPO system operating far below threshold.

We anticipate that with advancements in quantum optics techniques, it will be possible to generate two-mode SSCS with a large photon number and a quadrature squeezing parameter exceeding the currently reported value of -15 dB \cite{Vahlbruch16}. This paves the way for significant advancements in quantum optics and quantum information science, facilitated by the powerful capabilities of the $\Pi_{\rm N}$ nonclassicality indicator.




\begin{acknowledgments}
This work was supported by NRF-2022M3K4A109 4781, NRF-2019R1A2C2009974, and NRF-2022K1A3A1 A31091222.
\end{acknowledgments}

\appendix
\section{Two-mode quantum Langevin equation \label{sec:QLE}}
The quantum Langevin equations of motion for the slowly-varying intra-cavity operators  
$\tilde{a}_s$ and $\tilde{a}_i$ in the interaction picture with linear approximations \cite{Lee24,Collett84,Carlos22} can be written as follows:
\begin{equation}\label{E1E2}
		\frac{\partial \tilde{\boldsymbol{a}}}{\partial t}=\boldsymbol{\mathcal{L}}_0 \tilde{\boldsymbol{a}}+\sqrt{\gamma_1} \tilde{\boldsymbol{b}}_{\rm in}+\boldsymbol{\mathcal{A}}+\sqrt{\gamma_2}\tilde{\boldsymbol{c}}_{\rm in},
	\end{equation}
where $\tilde{\boldsymbol{a}}=\left(\tilde{a}_s,\tilde{a}_s^\dagger,\tilde{a}_i,\tilde{a}_i^\dagger\right)^T$ is the four-dimensional vector for intra-cavity bosonic slowly-varying operators, $\tilde{\boldsymbol{b}}_{\rm in}$ and $\tilde{\boldsymbol{c}}_{\rm in}$ are the four-dimensional vector for slowly-varying input operators associated to the vacuum operators, $\boldsymbol{\mathcal{A}}=\left(\mathcal{A}_s,\mathcal{A}_s^*,\mathcal{A}_i,\mathcal{A}_i^*\right)^T$, and 
	\begin{equation}
		\boldsymbol{\mathcal{L}}_0=
		-\frac{1}{2}\begin{pmatrix}
			\gamma_{12} & 0 & 0 & \gamma_s \\
			0 & \gamma_{12} & \gamma_s^*  & 0\\
			0 & \gamma_s  & \gamma_{12} & 0\\
			\gamma_s^*  & 0 & 0 & \gamma_{12}
		\end{pmatrix}.
	\end{equation}
Here, we used shorthand notations; $\tilde{o}(t) = \hat{o}e^{i\omega_jt}$, $o =\{a_j, b_{j,{\rm in}}, c_{j,{\rm in}}\}$, $j \in \{s,i\}$, {$\gamma_{12}=\gamma_1+\gamma_2$}, and $\mathcal{A}_j$ are defined in Sec.~\ref{sec:IH}. In addition, we note that the input operators $\hat{o}_{j,{\rm in}}$ in Eqs.~(\ref{E1E2}) are the input vacuum operators to the signal and idler modes with the dimension of Hz$^{1/2}$. They enter the OPO cavity through the mirrors MI and MO in Fig.~\ref{Fig0} from the outside vacuum with continuous spectra. Equation~(\ref{E1E2}) is derived by the unitary transformation of the intra-cavity Langevin equation for $\hat{a}_s$ and $\hat{a}_i$ with the intra-cavity field Hamiltonian $\hat{\mathcal{H}}_{\rm si}$ in Eq.~(\ref{Hamil}). Equation~(\ref{E1E2}) is valid only in the below-threshold region where $|\gamma_s'|<1$. Otherwise, $\boldsymbol{\mathcal{L}}_0$ possess eigenvalue with a non-negative real part \cite{Countinho05,Golubeva08,Ruiz18,Carlos22}. 

Now, to solve the quantum Langevin equation in Eq.~(\ref{E1E2}), we transform the slowly-varying operators into the frequency domain by the Fourier transform around the carrier frequency $\omega_j$, such that
\begin{align}
	\tilde{o}(\omega_j+\Omega_j)&=\frac{1}{\sqrt{2\pi}}\int_{-\infty}^\infty \tilde{o}(t) e^{i\Omega_j t} dt\nonumber\\
	& = \frac{1}{\sqrt{2\pi}}\int_{-\infty}^\infty \hat{o}(t) e^{i(\omega_j+\Omega_j) t} dt.\label{E6} 
\end{align}
We note here that the intra-cavity operators $\{\tilde{a}_s(\omega),\tilde{a}_i(\omega)\}$ have the physical dimension of Hz$^{-1}$, and after the Fourier transform, they become dimensionless in the time domain. Then, the solutions of the slowly-varying intra-cavity operators at the signal and idler frequencies can be obtained analytically and summarized as
\begin{widetext}
	\begin{align}
		\tilde{a}_j(\omega_j+\Omega_j) =& 
		\frac{2}{(\gamma_{12}-2i\Omega)^2-|\gamma_s|^2}\Big[\left(\gamma_{12}-2i\Omega_j\right)\left(\sqrt{\gamma_1}\tilde{b}_{j,{\rm in}}(\omega_j+\Omega_j)+\sqrt{\gamma_2}\tilde{c}_{j,{\rm in}}(\omega_j+\Omega_j)\right)  \nonumber \\
		&-\gamma_s\left(\sqrt{\gamma_1}\tilde{b}_{l,{\rm in}}^\dagger(\omega_l-\Omega_j)+\sqrt{\gamma_2}\tilde{c}_{l,{\rm in}}^\dagger(\omega_l-\Omega_j)\right) +\sqrt{2\pi}\delta(\Omega_j)\left(\left(\gamma_{12}-2i\Omega_j\right)\mathcal{A}_j-\gamma_s {\mathcal{A}_l}^*\right) \Big], \label{E7}
	\end{align}
\end{widetext}
where $\{j,l\} = \{s,i\}$ and $\delta(\cdot)$ is the Dirac delta function. 
As one can see, the solutions for $\tilde{a}_s(\omega_s+\Omega_s)$ and $\tilde{a}_i(\omega_i+\Omega_i)$ in Eq.~(\ref{E7}) are coupled with the vacuum input operators $\tilde{b}_{j,{\rm in}}$ and $\tilde{c}_{j,{\rm in}}$ as well as the coherent amplitudes $\mathcal{A}_s$ and $\mathcal{A}_i$. We point out here that the Fourier frequencies $\Omega_s$ and $\Omega_i$ in Eq.~(\ref{E7}) are defined within the range of $-\omega_{\rm FSR}/2 \le \Omega_j \le \omega_{\rm FSR}/2$. 
However, in the calculations below, for simplicity, we shall extend the range of Fourier frequencies up to $\pm\infty$ by considering the narrow linewidth condition of $\Delta\omega_c = \omega_{\rm FSR}/\mathcal{F} \ll \omega_{\rm FSR}$, where $\mathcal{F}$ is the finesse of the OPO cavity, which can be usually satisfied in quantum optics experiments.

Next, we can use the solutions for  $\tilde{a}_s(\omega_s+\Omega_s)$ and $\tilde{a}_i(\omega_i+\Omega_i)$ in Eq.~(\ref{E7}) to derive the expressions for slowly varying symmetric and antisymmetric operators $\tilde{A}_1(\Omega)$ and  $\tilde{A}_2(\Omega)$ in the frequency domain, those are the Fourier transformed operators of $\hat{A}_1$ and $\hat{A}_2$ in Eq.~(\ref{A1A2}). We obtained the final expressions as 
\begin{widetext}
	\begin{subequations}\label{E11E12}
		\begin{align}
			\tilde{A}_1(\Omega) =&
			\frac{2}{(\gamma_{12}-2i\Omega)^2-|\gamma_s|^2}\Big[(\gamma_{12}-2i\Omega)(\sqrt{\gamma_1}\tilde{B}_{1,{\rm in}}(\Omega)+\sqrt{\gamma_2}\tilde{C}_{1,{\rm in}}(\Omega))-\gamma_s(\sqrt{\gamma_1}\tilde{B}_{1,{\rm in}}^\dagger(-\Omega) \nonumber \\ &+ \sqrt{\gamma_2}\tilde{C}_{1,{\rm in}}^\dagger(-\Omega)) 
			+\sqrt{\pi}\delta(\Omega)((\gamma_{12}-2i\Omega)(\mathcal{A}_s+\mathcal{A}_i)-\gamma_s(\mathcal{A}^*_s+\mathcal{A}^*_i))\Big], \label{E11} \\
			\tilde{A}_2(\Omega) =&
			\frac{2}{(\gamma_{12}-2i\Omega)^2-|\gamma_s|^2}\Big[(\gamma_{12}-2i\Omega)(\sqrt{\gamma_1}\tilde{B}_{2,{\rm in}}(\Omega) +\sqrt{\gamma_2}\tilde{C}_{2,{\rm in}}(\Omega)) +\gamma_s(\sqrt{\gamma_1}\tilde{B}_{2,{\rm in}}^\dagger(-\Omega)\nonumber \\&+\sqrt{\gamma_2}\tilde{C}_{2,{\rm in}}^\dagger(-\Omega)) 
			+\sqrt{\pi}\delta(\Omega)((\gamma_{12}-2i\Omega)(\mathcal{A}_s-\mathcal{A}_i)+\gamma_s(\mathcal{A}^*_s-\mathcal{A}^*_i))\Big], \label{E12}
		\end{align}
	\end{subequations}
\end{widetext}
where $\Omega$ is the Fourier frequency, $\tilde{B}_{k,{\rm in}} = \frac{1}{\sqrt{2}}(\tilde{b}_{s,{\rm in}} \pm \tilde{b}_{i,{\rm in}})$, $\tilde{C}_{k,{\rm in}} = \frac{1}{\sqrt{2}}(\tilde{c}_{s,{\rm in}} \pm \tilde{c}_{i,{\rm in}})$, and $k \in \{1,2\}$. 

We stress that the input operators $\tilde{B}_{k,{\rm in}}(\Omega)$  and $\tilde{C}_{k,{\rm in}}(\Omega)$ in Eqs.~(\ref{E11E12})  are vacuum operators that their mean values are zero \cite{Carlos22}, i.e., $\langle \tilde{B}_{k,{\rm in}}\rangle = \langle \tilde{C}_{k,{\rm in}}\rangle = 0$. Moreover, since $\tilde{A}_1(\Omega)$ and $\tilde{A}_2(\Omega)$ are mutually independent, we can now investigate the properties of the two-mode squeezed state separately for $\tilde{A}_1(\Omega)$ and $\tilde{A}_2(\Omega)$.            

Next, to investigate the properties of the two-mode squeezed state outside the OPO cavity, we need to find the symmetric and antisymmetric output operators $\tilde{C}_{1,{\rm out}}(\Omega)$ and $\tilde{C}_{2,{\rm out}}(\Omega)$  from the corresponding intra-cavity operators $\tilde{A}_1$ and $\tilde{A}_2$ given in Eq.~(\ref{E11E12}) by using the input-output operator relations \cite{Collett84,Carlos22}
\begin{equation}
	\tilde{C}_{k,{\rm out}}(\Omega) = \sqrt{\gamma_2}\tilde{A}_k(\Omega)-\tilde{C}_{k,{\rm in}}(\Omega). \label{E15}
\end{equation}
Note here that the first term in the right-hand side of Eq.~(\ref{E15}) is the transmitted intra-cavity operator through the output mirror MO of the OPO cavity, and the second one is the input operator reflecting off the output mirror. One can derive the final analytical expressions for $\tilde{C}_{1,{\rm out}}(\Omega)$ and $\tilde{C}_{2,{\rm out}}(\Omega)$ as 
\begin{widetext}
	\begin{subequations}\label {Cout}
		\begin{align}
			\tilde{C}_{1,{\rm out}}(\Omega) =&
			\frac{1}{(\gamma_{12}-2i\Omega)^2-|\gamma_s|^2}\Big[\left(\gamma_2^2-\left(\gamma_1-2i\Omega\right)^2+|\gamma_s|^2\right)\tilde{C}_{1,{\rm in}}(\Omega) \nonumber \\
			&+{2}\sqrt{\gamma_1\gamma_2}\left(\gamma_{12}-2i\Omega\right)\tilde{B}_{1,{\rm in}}(\Omega)
			-{2}\gamma_s\gamma_2\tilde{C}_{1,{\rm in}}^\dagger(-\Omega)-{2}\gamma_s\sqrt{\gamma_1\gamma_2}\tilde{B}_{1,{\rm in}}^\dagger(-\Omega) \nonumber \\
			&+\sqrt{4\pi\gamma_2}\delta(\Omega)\left(\left(\gamma_{12}-2i\Omega\right)(\mathcal{A}_s+\mathcal{A}_i)-\gamma_s(\mathcal{A}^*_s+\mathcal{A}^*_i)\right) \Big], \label{Couta}\\
			\tilde{C}_{2,{\rm out}}(\Omega) =&
			\frac{1}{(\gamma_{12}-2i\Omega)^2-|\gamma_s|^2}\Big[\left(\gamma_2^2-\left(\gamma_1-2i\Omega\right)^2+|\gamma_s|^2\right)\tilde{C}_{2,{\rm in}}(\Omega) \nonumber \\
			&+{2}\sqrt{\gamma_1\gamma_2}\left(\gamma_{12}-2i\Omega\right)\tilde{B}_{2,{\rm in}}(\Omega)
			+{2}\gamma_s\gamma_2\tilde{C}_{2,{\rm in}}^\dagger(-\Omega)+{2}\gamma_s\sqrt{\gamma_1\gamma_2}\tilde{B}_{2,{\rm in}}^\dagger(-\Omega) \nonumber \\
			&+\sqrt{4\pi\gamma_2}\delta(\Omega)\left(\left(\gamma_{12}-2i\Omega\right)(\mathcal{A}_s-\mathcal{A}_i)+\gamma_s(\mathcal{A}^*_s-\mathcal{A}^*_i)\right) \Big].\label{Coutb}
		\end{align}
	\end{subequations}
	
\end{widetext}
As a last step, to derive the solutions for the symmetric and antisymmetric position  
$\tilde{X}_{k,{\rm out}}(\Omega)$ and momentum $\tilde{P}_{k,{\rm out}}(\Omega)$ quadrature operators of the output field from Eq.~(\ref{Cout}), we will use a simple way introduced in Ref.~\cite{Collett84}. Specifically, we define rotated quadrature operators $\tilde{X}_{k,{\rm out}}^{\phi_p}(\Omega)$ and $\tilde{P}_{k,{\rm out}}^{\phi_p}(\Omega)$ as the symmetric and antisymmetric position and momentum quadrature operators in the rotated frame. They are rotated by an squeezing angle of  $\phi_p/2$ from the unrotated frame. In this rotated frame, the symmetric position quadrature operator $\tilde{X}_{1,{\rm out}}^{\phi_p}$ lies along the semi-major squeezing axis, while the symmetric momentum quadrature operator $\tilde{P}_{1,{\rm out}}^{\phi_p}$ lies along the semi-minor antisqueezing axis. The antisymmetric quadrature operators exhibit opposite squeezing behavior. Finally, the rotated quadrature operators $\tilde{X}_{k,{\rm out}}^{\phi_p}(\Omega)$ and $\tilde{P}_{k,{\rm out}}^{\phi_p}(\Omega)$ can be written as \cite{Collett84} 
\begin{subequations}\label{E16}
	\begin{align}
		\tilde{X}_{k,{\rm out}}^{\phi_p}(\Omega)&=\frac{1}{\sqrt{2}}\left(\tilde{C}_{k,{\rm out}}(\Omega)e^{-i\phi_p/2}+\tilde{C}^\dagger_{k,{\rm out}}(\Omega)e^{i\phi_p/2}\right),  \label{E16a} \\ \tilde{P}_{k,{\rm out}}^{\phi_p}(\Omega)&=\frac{1}{\sqrt{2}i}\left(\tilde{C}_{k,{\rm out}}(\Omega)e^{-i\phi_p/2}-\tilde{C}^\dagger_{k,{\rm out}}(\Omega)e^{i\phi_p/2}\right). \label{E16b}
	\end{align} 
\end{subequations}

Now, we can easily find the mean values of the position $\langle\bar{X}_{k,{\rm out}}^{\phi_p}\rangle$ and momentum $\langle\bar{P}_{k,{\rm out}}^{\phi_p}\rangle$ quadrature operators after the integration over $\Omega$ from Eqs.~(\ref{Cout}) and (\ref{E16}). These are, in fact, the displacements $d X_{k,{\rm out}}^{\phi_p}$ and $dP_{k,{\rm out}}^{\phi_p}$, respectively, and they depend both on the injection rates $\mathcal{A}_j$ and squeezing rate $\gamma_s$ as expected in our BSM as 
\begin{subequations}\label{E17}
	\begin{align}
		dX_{k,{\rm out}}^{\phi_p} &= \langle\bar{X}_{k,{\rm out}}^{\phi_p}\rangle = \frac{2\sqrt{2\pi\gamma_2}}{\gamma_{12}\pm |\gamma_s|}\mathrm{Re}\lbrace \left(\mathcal{A}_s\pm\mathcal{A}_i\right)e^{-i\phi_p/2} \rbrace, \label{E17a}\\
		dP_{k,{\rm out}}^{\phi_p} &= \langle\bar{P}_{k,{\rm out}}^{\phi_p}\rangle = \frac{2\sqrt{2\pi\gamma_2}}{\gamma_{12}\mp|\gamma_s|}\mathrm{Im}\lbrace \left(\mathcal{A}_s \pm \mathcal{A}_i\right)e^{-i\phi_p/2}\rbrace, \label{E17b}
	\end{align}
\end{subequations}
where combinations for $k$ and signs $\pm \,\&\, \pm$ in Eq.~(\ref{E17a}) and $\mp \,\&\, \pm$ in Eq.~(\ref{E17b}) should be applied for $k = 1$ to the upper sign pair, and $k = 2$ to the lower sign pair, respectively. We want to point out here that in the rotated frame, although the squeezing variance is not dependent on the squeezing angle $\phi_p$ \cite{Collett84}, the expectation values of the symmetric and antisymmetric position and momentum quadrature operators, and equivalently the displacements associated with them, depend on the phase angle $\phi_p$, as can be seen in Eqs.~(\ref{E17}). We examine these features when we analyze the Wigner functions in Fig.~\ref{FigV2} below.

From Eqs.~(\ref{E16}) and (\ref{E17}), we can also calculate directly the two-mode squeezing spectrum $S_{\rm out}(\Omega)$ as well as the two-mode variance $V_{\rm out}(\Omega)$, as summarized in Eqs.~(\ref{E18}) and (\ref{E19}) below. Note here that, since the constant terms involving $\mathcal{A}_j$ in Eqs.~(\ref{E11E12}) are canceled out in their covariances $\langle \tilde{C}_{k,{\rm our}}(\Omega), \tilde{C}_{k,{\rm our}}(\Omega')\rangle$ and $\langle \tilde{C}_{k,{\rm our}}^\dagger(\Omega), \tilde{C}_{k,{\rm our}}(\Omega')\rangle$, the analytical solutions for  $S_{\rm out}(\Omega)$ and $V_{\rm out}(\Omega)$ depend only on the squeezing parameter $|\gamma_s|$.

\begin{figure}[t]
	\centering
	\includegraphics[width=8 cm]{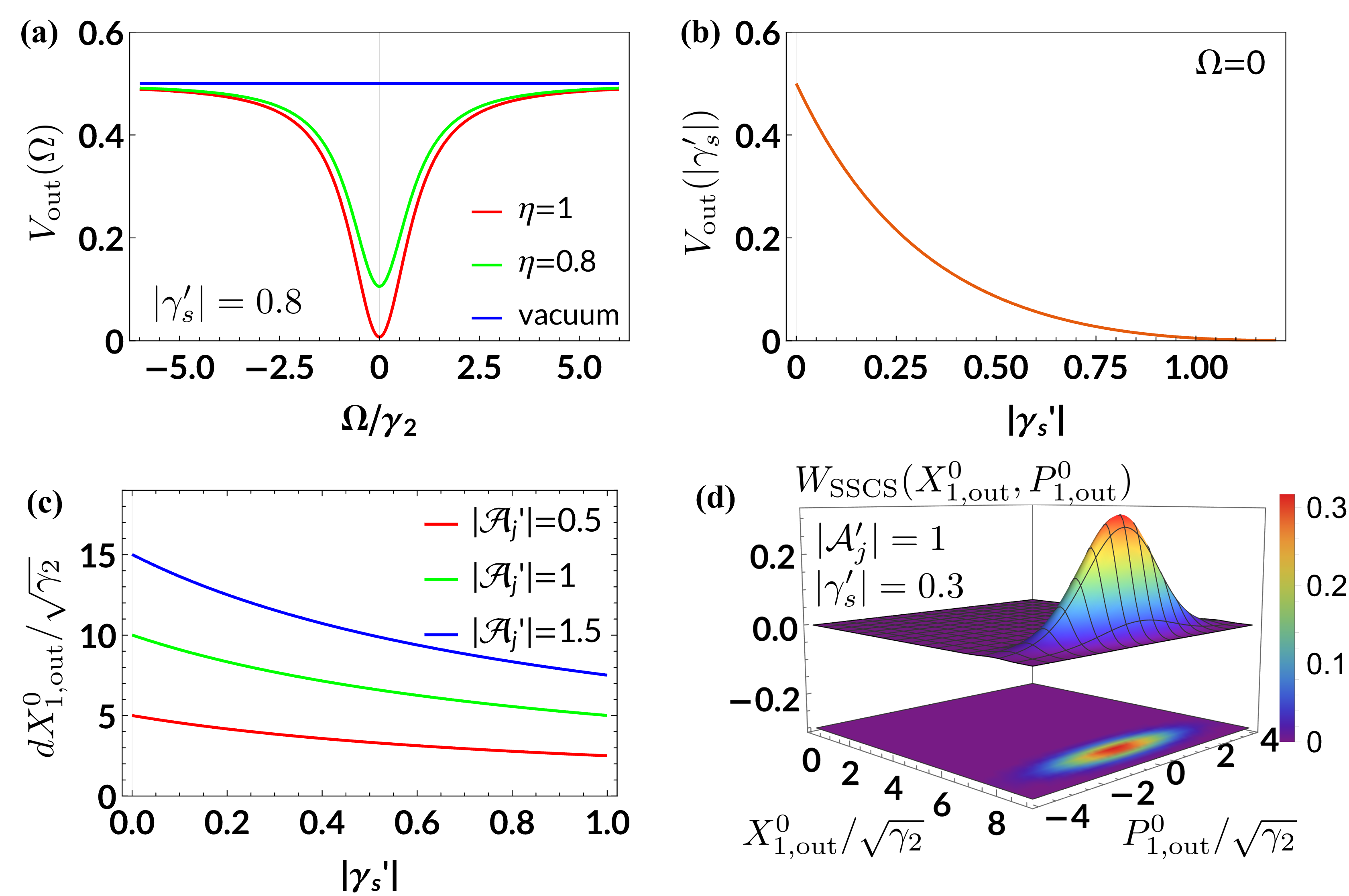}
	\caption{(a) $V_{\rm out}$ versus $\Omega$ for vacuum (blue solid) and two-mode SSCS with $\eta=1$ (red solid) and $\eta = 0.8$ (green solid) at $|\gamma_s'| = 0.8$. (b) $V_{\rm out}$ versus $|\gamma_s'|$. (c) $dX_{k,{\rm out}}^{\phi_p} = \langle \bar{X}_{1,{\rm out}}^{\phi_p}\rangle$ versus $|\gamma_s'|$ for three different injection rates: $|\mathcal{A}_j'| = \{0.5 \,({\rm red\, solid}), 1.0 \,({\rm blue\, solid}), 1.50 \,({\rm red\, solid})\}$. (d) Wigner function $W_{\rm SSCS}(X_{1,{\rm out}}^0,P_{1,{\rm out}}^0)$ of the state $|\psi_1(\tau)\rangle$ in Eq.~(\ref{taus1}) at steady state with  $|\gamma_s'| = 0.3$ and $|\mathcal{A}_j'| = 1$, i.e., $dX_{1,{\rm out}}^0 = 7.7$ from Eq.~(\ref{E17}). In (b), (c), and (d), we used the parameters of $\eta = 1$, $\Omega = 0$, and $\gamma_1 = 0.002\gamma_2$.} \label{FigV2}
\end{figure}

Finally, we obtained the analytical expression for two-mode squeezing spectrum $S_{\rm out}(\Omega)$, which is equal to the normally ordered covariance associated with either the total position quadrature operator $\tilde{X}_{1,{\rm out}}^{\phi_p}(\Omega)$ or the relative momentum quadrature operator $\tilde{P}_{2,\rm out}^{\phi_p}(\Omega)$ \cite{Collett84,Carlos22} as
\begin{align}
	S_{\rm out}(\Omega) &= \int d\Omega' \langle : \tilde{X}_{1,{\rm out}}^{\phi_p}(\Omega),\tilde{X}_{1,{\rm out}}^{\phi_p}(\Omega') : \rangle\nonumber\\ 
	&= \int d\Omega' \langle : \tilde{P}_{2,{\rm out}}^{\phi_p}(\Omega),\hat{P}_{2,{\rm out}}^{\phi_p}(\Omega') : \rangle \nonumber \\
	&= -\frac{2\eta |\gamma_s|\gamma_2}{\left(\gamma_{12} + |\gamma_s|\right)^2+4\Omega^2}, \label{E18} 
\end{align}
where $\langle:\cdot:\rangle$ stands for the normal ordering and $\eta$ is the homodyne detection efficiency. As a result, the two-mode squeezing variance $V_{\rm out}(\Omega)$ can also be obtained by simply adding the two-mode vacuum variance of $\frac{1}{2}$ to the squeezing spectrum as 
\begin{equation}
	V_{\rm out}(\Omega) =  \frac{1}{2} + S_{\rm out}(\Omega)  = \frac{1}{2} -\frac{2\eta |\gamma_s|\gamma_2}{\left(\gamma_{12} + |\gamma_s|\right)^2+4\Omega^2}.\label{E19} 
\end{equation}

As clearly seen in Eqs.~(\ref{E18}) and (\ref{E19}), the squeezing spectrum $S_{\rm out}(\Omega)$ and the variance $V_{\rm out}(\Omega)$ are independent of the bichromatic injection rates $\mathcal{A}_s$ and $\mathcal{A}_i$, but depend only on the squeezing parameter $|\gamma_s|$. On the other hand, the mean values $\langle\bar{X}_{k,{\rm out}}^{\phi_p}\rangle$ and $\langle\bar{P}_{k,{\rm out}}^{\phi_p}\rangle$ in Eqs.~(\ref{E17}) depend on both $|\gamma_s|$ and $\mathcal{A}_j$. Figure~\ref{FigV2} summarizes the various Gaussian properties of the SSCS graphically. 

\section{Calculation of nonclassical properties}\label{sec:AppB}

To analyze the quantum correlations of the state, we define a second-order moment (SOM) function as
\begin{align}
	\Gamma &= \langle \hat{t}(\omega),\hat{u}(\omega'),\hat{v}(\omega''),\hat{w}(\omega''')\rangle \nonumber \\
	&\equiv\iiiint_{-\infty}^\infty d\omega d\omega' d\omega'' d\omega'''\left\langle \hat{t}(\omega)\hat{u}(\omega')\hat{v}(\omega'')\hat{w}(\omega'''\right\rangle.\label{S58}
\end{align}
In addition, for two commuting number operators $\hat{N}_j=\hat{a}_j^\dagger(\omega)\hat{a}_j(\omega)$, $j\in\{s,i\}$, and by rearranging the Cauchy-Schwarz inequality, 
we define a nonclassical measure $\Pi_{\rm A,in}$ (Eq.~(\ref{E23})) as the measure of the nonclassicality of the intra-cavity state \cite{Agarwal88} as 
\begin{equation}\label{S60}
	\Pi_{\rm A,in} = \frac{\sqrt{\langle: \hat{N}_s^2:\rangle \langle: \hat{N}_i^2:\rangle}}{|\langle \hat{N}_s\hat{N}_i \rangle|}.
\end{equation}

Now, to evaluate the nonclassical measure $\Pi_{\rm A,in}$  analytically, we need to calculate the SOM function defined in Eq.~(\ref{S58}) by using the expressions of slowly-varying intra-cavity signal and idler operators $\tilde{a}_s(\omega)$ and $\tilde{a}_i(\omega)$  in Eq.~(\ref{E7}). Then, we can calculate separately each term in the nonclassical measure $\Pi_{\rm A,in}$ as follows
\begin{align}
	\langle :\hat{N}_s^2:\rangle &=\Gamma_s = \langle \tilde{a}_s^\dagger(\omega)\tilde{a}_s^\dagger(\omega')\tilde{a}_s(\omega'')\tilde{a}_s(\omega''')\rangle, \label{S61}\\
	\langle: \hat{N}_i^2:\rangle &=\Gamma_i = \langle \tilde{a}_i^\dagger(\omega)\tilde{a}_i^\dagger(\omega')\tilde{a}_i(\omega'')\tilde{a}_i(\omega''')\rangle, \label{S62}\\
	\langle \hat{N}_s\hat{N}_i \rangle &=\Gamma_{si} = \langle \tilde{a}_s^\dagger(\omega)\tilde{a}_s(\omega')\tilde{a}_i^\dagger(\omega'')\tilde{a}_i(\omega''')\rangle, \label{S63}
\end{align}
where $\Gamma_l, l\in\{s, i, si\}$, is the SOM function defined in Eq.~(\ref{S58}) with three different combinations of intra-cavity boson operators.

Note here that the slowly varying operators $\tilde{a}_j(\omega)$ in Eqs.~(\ref{S61}) to (\ref{S63}) are actually intra-cavity signal and idler operators with optical frequencies $\omega_j$, i.e.,  $\tilde{a}_{j}(\omega) \equiv \tilde{a}_{j}(\omega_{j}+\omega)$, $j \in \{s,i\}$, $\omega_j \gg \omega_{\rm FSR}$ and $|\omega| < \omega_{\rm FSR}$. Since we are working in the Heisenberg picture and quantum Langevin equations to study the quantum dynamics of the slowly varying intra-cavity operators $\tilde{a}_s(\omega)$ and $\tilde{a}_i(\omega)$, the SOM functions can be easily evaluated with respect to the two-mode vacuum state $|0,0\rangle$ after inserting Eq.~({\ref{E7}) into Eqs.~(\ref{S61}) and (\ref{S63}). 
	
Let's evaluate some particular term of $\Gamma_s$ in Eq.~(\ref{S61}). To do this, we need to calculate the expectation values of the product of four intra-cavity signal and idler operators. One can see that $\Gamma_s$ has  total $5^4$ separate terms of input operators and coherent seed amplitudes. Among them, due to the fact that input annihilation operators have zero eigenvalues for the two-mode vacuum state, expectation values of the terms that have any input annihilation operator at the rightmost place automatically vanishes, resulting in non-zero expectation values from only the terms that have even number of input operators, i.e., $\tilde{j}_k, j \in \{ b,c, b^\dagger, c^\dagger\}, k \in \{(s,{\rm in}), (i,{\rm in})\}$ from the orthogonality of the number states. Indeed, we have identified that there are only 15 terms in $\Gamma_s$ and $\Gamma_i$ as well as 17 terms in $\Gamma_{si}$ that have non-zero contributions, including four pairs of 2 isomorphic terms in $\Gamma_s$ and $\Gamma_i$ with four input operators. We show here only one typical example that has non-zero expectation value, in which one of the creation operator  $\{\tilde{b}_j^\dagger,\tilde{c}_j^\dagger \}$, $j \in \{s,i\}$, is placed at the rightmost place such as

\begin{align}
	\Gamma_{s1} &= \int_{-\infty}^\infty d^4\omega {\rm C}^4|\gamma_s|^2\gamma_2^2(\gamma_{12}-2i\omega')(\gamma_{12}+2i\omega'')\nonumber\\
	&\times\langle\tilde{c}_{i,{\rm in}}(-\omega)
	\tilde{c}_{i,{\rm in}}^\dagger(\omega'')\tilde{c}_{s,{\rm  in}}(\omega')\tilde{c}_{s,{\rm in}}^\dagger(-\omega''') \rangle,\label{S64}
\end{align}
where ${\rm C}^4 = C(\omega)C(\omega')C(\omega'')C(\omega''')$ and $C(\omega)={2}/{\left[\left(\gamma_{12}-2i\omega\right)^2-|\gamma_s|^2\right]}$.

The expectation value in Eq.~(\ref{S64}) can now be simplified by using the bosonic commutation relations such that $\langle \tilde{c}_{i,{\rm in}}(-\omega)\tilde{c}_{i,{\rm in}}^\dagger(\omega'')\tilde{c}_{s,{\rm in}}(\omega')\tilde{c}_{s,{\rm in}}^\dagger(-\omega''')\rangle=\delta(\omega+\omega'')\delta(\omega'+\omega''')$. Therefore, we can now evaluate simply the four-dimensional (4-D) integration in Eq.~(\ref{S64}), resulting in $\Gamma_{s1}$ in a very compact form as
	\begin{equation}
		\Gamma_{s1} = |\gamma_s|^2\gamma_2^2 \left(\frac{\pi}{\gamma_{12}^2-|\gamma_s|^2}\right)^2. \label{S67}
	\end{equation}
	
It is now straightforward to find the sum of total 15 non-zero terms for $\Gamma_s$ and $\Gamma_i$ and 17 terms for $\Gamma_{si}$ with the same way as discussed above to {obtain Eq.~(\ref{S67}).} After a lengthy calculation, even though we do not write every result of calculations here, because it takes too many pages and it is hard to gain physical insights from them, we write only the final results for $\Gamma_{s}$ and $\Gamma_i$ as well as $\Gamma_{si}$ for the intra-cavity signal and idler modes. The final expressions for $\Gamma_s$, $\Gamma_i$, and $\Gamma_{si}$ may be, respectively, summarized as
\begin{widetext}
	\begin{align}
		\Gamma_s&=\frac{\pi^2}{2}|\gamma_s|^4 \,{\rm Qs}^2
		+2\pi|\gamma_s|^2\,{\rm Qs}^3\left|{\rm As}\right|^2 + {\rm Qs}^4 \left|{\rm As}\right|^4, \label{S68}\\
		\Gamma_i&=\frac{\pi^2}{2}|\gamma_s|^4\,{\rm Qs}^2
		+2\pi|\gamma_s|^2\,{\rm Qs}^3\left|{\rm Ai}\right|^2 + {\rm Qs}^4 \left|{\rm Ai}\right|^4, \label{S69}\\
		\Gamma_{si}&=\frac{\pi^2}{4}|\gamma_s|^2\left(\gamma_{12}^2+|\gamma_s|^2\right){\rm Qs}^2
		-\pi\gamma_{12} \Re\{
		\gamma_s^*{\rm As Ai}\}{\rm Qs}^3 +\frac{\pi}{2}|\gamma_s|^2 \left(|{\rm As}|^2+|{\rm Ai}|^2\right) {\rm Qs}^3  + |{\rm As}|^2|{\rm Ai}|^2 {\rm Qs}^4, \label{S72}
	\end{align} 
	where 
	\begin{align}
		{\rm Qs} &= \frac{2}{\gamma_{12}^2-|\gamma_s|^2}, \nonumber \\
		{\rm As} &= \sqrt{2\pi}(\gamma_{12}\mathcal{A}_s-\gamma_s\mathcal{A}_i^*)= \sqrt{\gamma_1}\left(\gamma_{12}|\alpha_s|e^{i\phi_s}-|\gamma_s||\alpha_i|e^{i(\phi_p-\phi_i)}\right), \label{S70} \\
		{\rm Ai} &= \sqrt{2\pi}(\gamma_{12}\mathcal{A}_i-\gamma_s\mathcal{A}_s^*)= \sqrt{\gamma_1}\left(\gamma_{12}|\alpha_i|e^{i\phi_i}-|\gamma_s||\alpha_s|e^{i(\phi_p-\phi_s)}\right), \label{S71} 
	\end{align}
\end{widetext}
	and $\alpha_s = |\alpha_s|e^{i\phi_s}$ and $\alpha_i=|\alpha_i|e^{i\phi_i}$ are the coherent amplitudes of the signal and idler seed fields.	
	
Finally, by using the input-output relation, we can obtain the expression for the non-classical measure $\Pi_{\rm A, out}$ of the two-mode SSCS at the output of the OPO cavity following the exactly same procedure. After a lengthy calculation, we found a very simple result as 
\begin{equation}
		\Pi_{\rm A, out}= \frac{\sqrt{\Gamma_{s}^{\rm out}\Gamma_i^{\rm out}}}{|\Gamma_{si}^{\rm {out}}|}  = \Pi_{\rm A, in}, \label{S74}
\end{equation} 
where individual SOM functions are found to be
\begin{align}
		\Gamma_s^{\rm out} &= \langle\tilde{a}^\dagger_{s,{\rm out}}(\omega),\tilde{a}^\dagger_{s,{\rm out}}(\omega'),\tilde{a}_{s,{\rm out}}(\omega''),\tilde{a}_{s,{\rm out}}(\omega''')\rangle \nonumber \\
		&=\gamma_2^2 \Gamma_s, \label{S75}
\end{align}
\begin{align}
		\Gamma_i^{\rm out} &= \langle\tilde{a}^\dagger_{i,{\rm out}}(\omega),\tilde{a}^\dagger_{i,{\rm out}}(\omega'),\tilde{a}_{i,{\rm out}}(\omega''),\tilde{a}_{i,{\rm out}}(\omega''')\rangle \nonumber \\
		&=\gamma_2^2\Gamma_i, \label{S76}
\end{align}
\begin{align}
		\Gamma_{si}^{\rm out} &= \langle\tilde{a}^\dagger_{s,{\rm out}}(\omega),\tilde{a}_{s,{\rm out}}(\omega'),\tilde{a}^\dagger_{i,{\rm out}}(\omega''),\tilde{a}_{i,{\rm out}}(\omega''')\rangle \nonumber \\ &=\gamma_2^2\Gamma_{si}. \label{S77}
\end{align}

Similarly, we find the expression of $\Pi_{\rm L,out}$ by using the results given in Eqs.~(\ref{S68}), (\ref{S69}), and (\ref{S72}), and the  result is given in Eq.~(\ref{E28b}) in the main text.


\bibliography{exports}

\end{document}